\documentclass{aa}
\usepackage{times}
\usepackage{psfig}

\begin{document}

\thesaurus{02.08.1; 03.13.4; 09.11.1; 10(03.1; 11.1; 19.3)}

\title{3D self-consistent $N$-body barred models of the Milky~Way}
\subtitle{II. Gas dynamics}

\author{R. Fux}

\institute{Geneva Observatory, Ch. des Maillettes 51,
           CH-1290 Sauverny, Switzerland}

\offprints{R. Fux}
\mail{Roger.Fux@obs.unige.ch}

\date{Received 26 November 1998 / Accepted 9 February 1999}
\maketitle

\begin{abstract}
The gas dynamics in the Galactic disc is modeled by releasing an
initially axisymmetric SPH component in a completely self-consistent
and symmetry-free 3D $N$-body simulation of the Milky~Way in which the
stellar components display a COBE-like bar. The density centre of the
stellar bar wanders around the centre of mass and the resulting gas
flow is {\it asymmetric} and {\it non-stationary}, reproducing the HI
and CO $\ell-V$ diagrams only at specific times and thus suggesting a
transient nature of the observed inner gas kinematics.
\par The best matching models allow a new and coherent interpretation
of the main features standing out of the $\ell-V$ data within the bar
region. In particular, the $\ell-V$ traces of the prominent offset
dustlanes leading the bar major axis in early-type barred spirals can
be unambiguously identified, and the 3-kpc arm and its non-symmetric
galactocentric opposite counterarm are the inner prolongations of disc
spiral arms passing round the bar and joining the dustlanes at very
different galactocentric distances. Bania's clump~1 and 2, and another
velocity-elongated feature near $\ell= 5.5\degr$, are interpreted
as gas lumps crossing the dustlane shocks. The terminal velocity peaks
near $\ell=\pm 2.5\degr$ are produced by gas along the dustlanes and
not by the trace of the cusped $x_1$ orbit, which passes farther away
from the Galactic centre. According to these models and to related
geometrical constraints, the Galactic bar must have an inclination
angle of $25\degr\pm 4\degr$, a corotation radius of $4.0-4.5$~kpc
and a face-on axis ratio $b/a\approx 0.6$.
\end{abstract}

\section{Introduction}
%
The gas kinematics near the Galactic plane, mainly observed in HI
(Burton 1985; Kerr et al. 1986; Stark et al. 1992 and Hartmann \&
Burton 1997), $^{12}$CO (Dame et al. 1987; Oka et al. 1998b),
$^{13}$CO and CS (Bally et al. 1987), is probably among the best
tracer of the dynamical mass in the Galaxy. Under the assumption of
circular motion in an axisymmetric potential, the tangent point method
allows in principle to derive the rotation curve inside the solar
circle and to recover the spatial location of the spiral arms (see the
compilation of Vall\'ee 1995).
\par However, this approach is limited for at least two reasons
related to the data themselves. First, it is long known (Rougoor \&
Oort 1960) that the longitude-velocity ($\ell-V$) distribution of the
Galactic gas reveals substantial diffuse and feature-like emission in
the regions $(\ell>0,V<0)$ and $(\ell<0,V>0)$ close to the Galactic
centre which are forbidden to pure circular motion, the most popular
example being the 3-kpc arm. Second, bumps of about $7$~km\,s$^{-1}$
on the terminal velocity curves near the tangent points of the spiral
arms reflect the gravitational perturbation of these arms on the gas
flow and can propagate into distance errors of $\sim 1$~kpc along the
line of sight if modeled by circular motion (Burton 1992; Combes 1991).
\par Traditional interpretations of the ``forbidden'' velocities near
the centre involve explosion induced expansion (van der Kruit 1971;
van der Kruit et al. 1972; Cohen 1975; Oort 1977) or gas moving on
elliptical orbits in either a spiral (Shane~1972; Simonson \& Mader
1973) or a barred potential (Peters 1975; Cohen \& Few 1976; Liszt \&
Burton 1980; Gerhard \& Vietri 1986). In the last decade, the latter
interpretation received strong support by direct evidence of a large
scale stellar bar in the Milky~Way from near-IR surface photometry
(Blitz \& Spergel 1991; Dwek et al. 1995; Binney et al. 1997),
discrete source counts (Nakada et al. 1991 and Izumiura et al. 1994
for SiO masers, Weinberg 1992 and Nikolaev \& \mbox{Weinberg} 1997 for
AGB stars; Whitelock \& Catchpole 1992 for Miras variables; Blitz 1993
for globular clusters; Sevenster 1996 for OH/IR stars; Stanek et al.
1997 for red clump stars) and large microlensing optical depths
towards the bulge (Paczynski et al. 1994; Evans 1994; Zhao et al.
1996; Zhao \& Mao 1996; Gyuk~1999; see Gerhard 1996 and Kuijken 1996
for reviews).
\par Many hydrodynamical simulations (Roberts et al. 1979; van Albada
1985; Mulder \& Liem 1986; Athanassoula 1992; Englmaier \& Gerhard
1997) have shown that the gas flow in a rapidly rotating barred
potential is driven by strong shocks leading the bar major axis and
followed downstream by enhanced gas densities. In external early-type
(i.e. Sbc and earlier) barred spirals, these shocks are detected as
offset dustlanes with very large velocity changes in the associated
gas velocity field (e.g. Reynaud \& Downes 1998; Laine et al. 1999).
\par Binney et al.~(1991) have interpreted the gas kinematics near the
Galactic centre approximating the gas streamlines by a sequence of
closed $x_1$ and $x_2$ orbits (Contopoulos \& Mertz-anides 1977), but
such a model does not properly cares about the shocks. The hydro
simulations designed for the Milky~Way (Mulder \& Liem 1986; Wada et
al. 1994; Englmaier \& Gerhard 1999; Weiner \& Sellwood 1999) always
assume rigid barred potentials and bisymmetry with respect to the
Galactic centre, but some details in the HI and CO observations
clearly betray important asymmetries: (i) about 3/4 of the molecular
gas in the nuclear ring/disc lies at positive longitude, (ii) the
3-kpc arm and its far-side counterarm have very different absolute
velocities at $\ell=0$, and (iii) the positive and negative velocity
peaks of bright emission, at $(\ell,V/{\rm km\,s}^{-1})\approx
(3\degr,270)$ and $(-2\degr,-220)$ respectively, differ by roughly
$50$~km\,s$^{-1}$ in absolute velocity, and the emission of the
nuclear ring/disc is also shifted towards receding velocities.
Moreover, as an SBbc galaxy (Sackett 1997), the Milky~Way is also
expected to present prominent dustlanes, but their gaseous traces have
never been convincingly isolated in the HI and CO data.
\par This paper follows a first paper (Fux 1997, hereafter paper~I)
where several $N$-body barred models of the Galaxy have been built
from bar unstable axisymmetric initial conditions, constraining the
position of the observer relative to the bar with the COBE K-band data
corrected for extinction by dust. The best matching models suggested
an angle of $28\degr\pm 7\degr$ for the bar inclination and a
corotation radius of $4.3\pm 0.5$~kpc.~Here we take advantage of these
simulations to seek the most convenient initial conditions for much
larger simulations, including a gas component treated by the smooth
particle hydrodynamics (SPH) technique and with no imposed symmetries.
The resulting self-consistent gas flows are used to interpret the
observed gas kinematics near the Galactic plane, and more especially
within the bar region. Contrary to Weiner \& Sellwood~(1999), our
approach is based on the observed bright $\ell-V$ features, and not on
the mostly low density gas close to the terminal velocities, for which
Eulerian codes are more appropriate.
\par The structure of the paper is organised as follows: in Sect.~2 we
describe the main features appearing in the HI and CO $\ell-V$ diagrams
within the bar region. In Sect.~3 we expose the numerical $N$-body and
SPH code used to perform our simulations. In Sect.~4 and~5, we present
the initial conditions and the time evolution of these simulations. In
Sect.~6 we select some optimum models from the simulations and
interpret the observed gas kinematical features. In Sect.~7 we set
some new constraints on the bar parameters based on this
interpretation. Finally, Sect.~8 summarises our results. Throughout
the paper, we will adopt a solar galactocentric distance
$R_{\circ}=8$~kpc.

\section{Main observed features}
\begin{figure}[t!]
\centerline{\psfig{figure=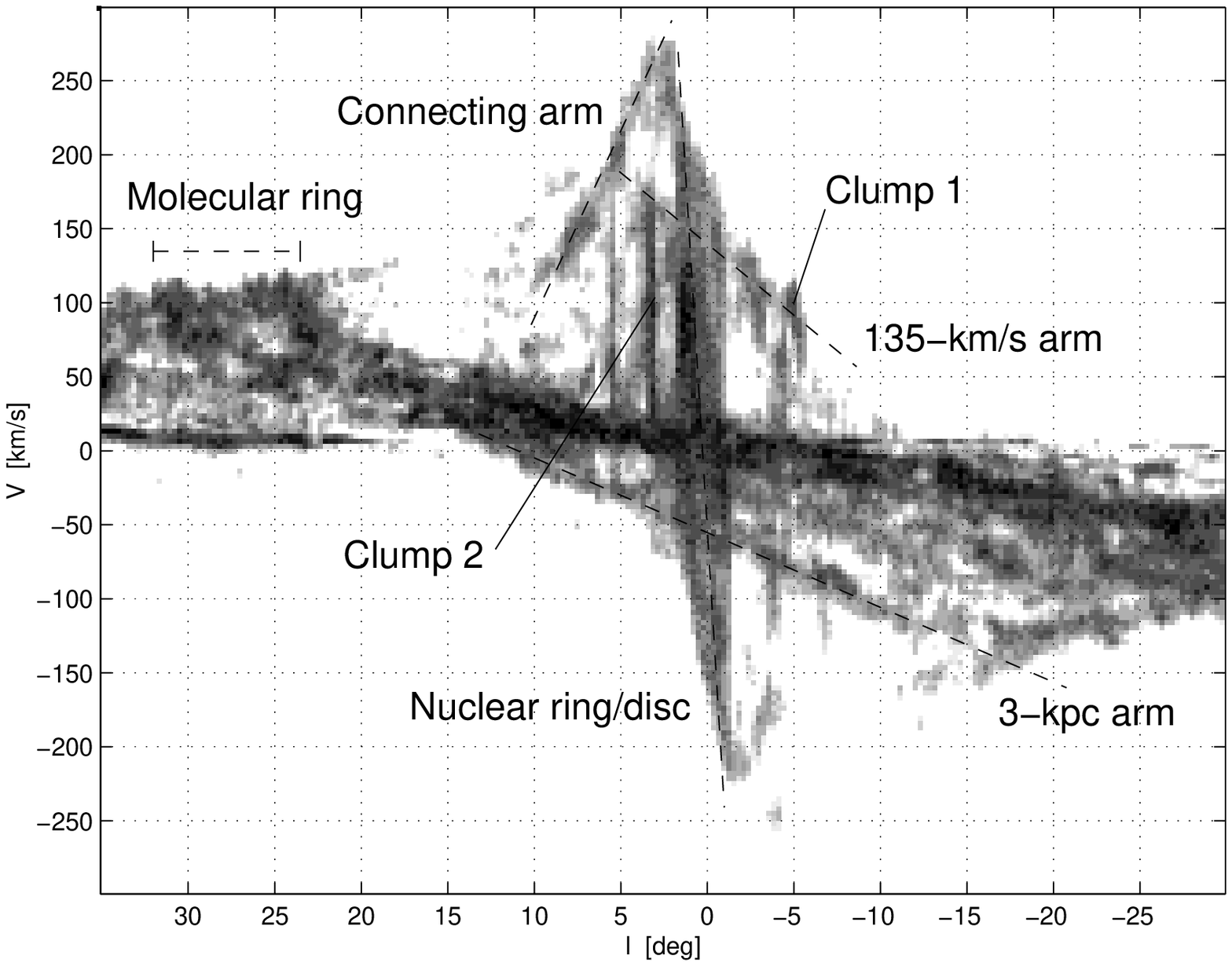,width=8.8cm}}
\caption{Grey scale longitude-velocity diagram of $^{12}$CO
         $J=1\rightarrow 0$ emission within
         $-30\degr\leq \ell \leq 35\degr$ and over a $4\degr$ latitude
         strip centred on the Galactic plane (Dame et al. 1999), with
         the dominant features indicated by dashed lines and
         Bania's~(1977) molecular clumps.}
\label{lv}
\caption{(right column) Longitude-velocity diagrams of HI $21$~cm
         (left) and $^{12}$CO $J=1\rightarrow 0$ (right) emission as a
         function of latitude in the range $|b|\leq 1.5\degr$. The HI
         data are from Hartmann \& Burton~(1997), Burton \&
         Liszt~(1978) and Kerr et al.~(1986), and the CO data from
         Dame et al.~(1987).}
\label{lvb}
\end{figure}

%
In the region $|\ell|\la 30\degr$ several features stand out from the
observed longitude-velocity distribution of both atomic and molecular
gas (see Figs.~\ref{lv} and~\ref{lvb}). The nomenclature of these
features principally refers to Rougoor~(1964) and we omit here the
historical but confusing ``expanding'' qualification, since local
radial motions do not necessarily imply a net flux when averaged over
azimuth. Other more exhaustive inventories of observed $\ell-V$
features can be found in van der Kruit~(1970) and Cohen~(1975) for the
HI, and in Bania~(1977), Bally et al.~(1988) and the review of
Combes~(1991) for the CO.
\begin{itemize}
\item {\it The 3-kpc arm}. This arm is the largest apparent feature,
covering more than $35\degr$ in Galactic longitude, and with a
``forbidden'' radial velocity of $-53$~km\,s$^{-1}$ at $\ell=0$. It
was discovered by van Woerden et al.~(1957) and its name is related to
the location of its tangent point near $\ell=-22\degr$, corresponding
to $R=3$ kpc (at this time the galactocentric distance of the Sun was
assumed to be $8.2$~kpc). The same authors also noticed that this arm
must lie in front of the Galactic centre because of absorbing the
continuum spectrum of radio sources close to the centre (see the
HI $\ell-V$ diagram at $b=0$ in Fig.~\ref{lvb}), which is consistent
with its rather large latitudinal width. The 3-kpc arm cannot be
represented by a uniformly rotating and expanding circular arc over
its whole longitude range (Burke \& Tuve 1964). It was successfully
modeled by Mulder \& Liem~(1986) as a stationary density wave in a
rotating barred potential.
\item {\it The 135-km\,s$^{-1}$ arm}. This feature shows no absorption
lines against Galactic centre radio sources and is suspected to be the
far-side counterpart of the 3-kpc arm (e.g. Oort 1977). However it
crosses the zero longitude axis at 135~km\,s$^{-1}$ (hence his name),
i.e. more than twice the absolute velocity of the 3-kpc arm, which is
among the strongest evidence for an asymmetric spiral structure in the
central few kpc (velocity asymmetries at $\ell=0$ cannot result from
perspective effects). The 135-km\,s$^{-1}$ arm extends down to
$\ell\approx -5\degr$, where its radial velocity still reaches about
$100$~km\,s$^{-1}$, and lies slightly above the Galactic plane, at
$b\approx 0.5\degr$. It also involves less HI mass than the 3-kpc arm
and is more lumpy.
\item {\it The connecting arm}. This is a very rarely discussed
feature in the recent literature, despite its substantial brightness.
It was probably mentioned the first time by Rougoor \& Oort~(1960).
The origin of its name comes from the fact that it seems to link the
nuclear ring/disc to the outer disc. It becomes easily identifiable in
the HI and $^{12}$CO data near $\ell=10\degr$ and
$V=110$~km\,s$^{-1}$, where it is located far below the plane at
$b\approx -1\degr$, and passes through the peak of the positive
terminal velocity curve. In Rougoor's~(1964) description, this feature
is doubtfully combined with another distinct feature extending at
roughly constant velocity out to $\ell\approx 18\degr$. The connecting
arm was soon recognized as a very inclined spiral arm, i.e. with a
high pitch angle, although its situation in front or behind the
Galactic centre remained unclear (Rougoor 1964; Cohen \& Davies 1976).
Kerr~(1967) adopted the latter alternative and interpreted this arm as
part of a central gaseous bar.
\item {\it The nuclear ring/disc}, or {\it central molecular zone}.
Concerns the offcentred dense molecular gas located within
$-1\degr\la \ell \la 1.5\degr$, i.e. somewhat inside the positive and
negative peaks of the terminal velocity curves (see for instance the
high resolution CO $\ell-V$ maps in Oort 1977; Liszt \& Burton 1978;
Bally et al. 1988; Combes 1991; Morris \& Serabyn 1996 and Oka et al.
1998b). The parallelogram bounding this structure is also called
``expanding molecular ring'' owing to its original interpretation
(Scoville 1972; Kaifu et al. 1972), and more recently ``180-pc ring''
according to its size. The molecular complex enclosed by this ring
forms a kind of disc rotating with a maximum velocity of order
$100$~km\,s$^{-1}$ and is speculated to form two mini spirals (Sofue
1995). Almost the entire CS emission comes from this disc. Binney et
al.~(1991) have interpreted the parallelogram and the disc
respectively as gas on the cusped $x_1$ orbit and on $x_2$ orbits in
the bar potential (see Sect.~\ref{bin}). An extensive review on this
pattern is given by Morris \& Serabyn~(1996).
\item {\it The molecular ring}. Meant to represent a ring-like gas
concentration with a radius of roughly $4$~kpc, the spatial
distribution of this structure is in fact poorly known and could as
well involve imbricated spiral arms. Close to the positive terminal
velocity curve, it forms two branches with tangent points at
$\ell\approx 25\degr$ and $30\degr$, corresponding to
$R\approx 3.4$~kpc and $4$~kpc. The molecular ring probably compares
to the inner (pseudo-) rings seen in external spiral galaxies (Buta
1996).\vspace*{-.05cm}
\end{itemize}
\begin{figure}[p!]
\centerline{\psfig{figure=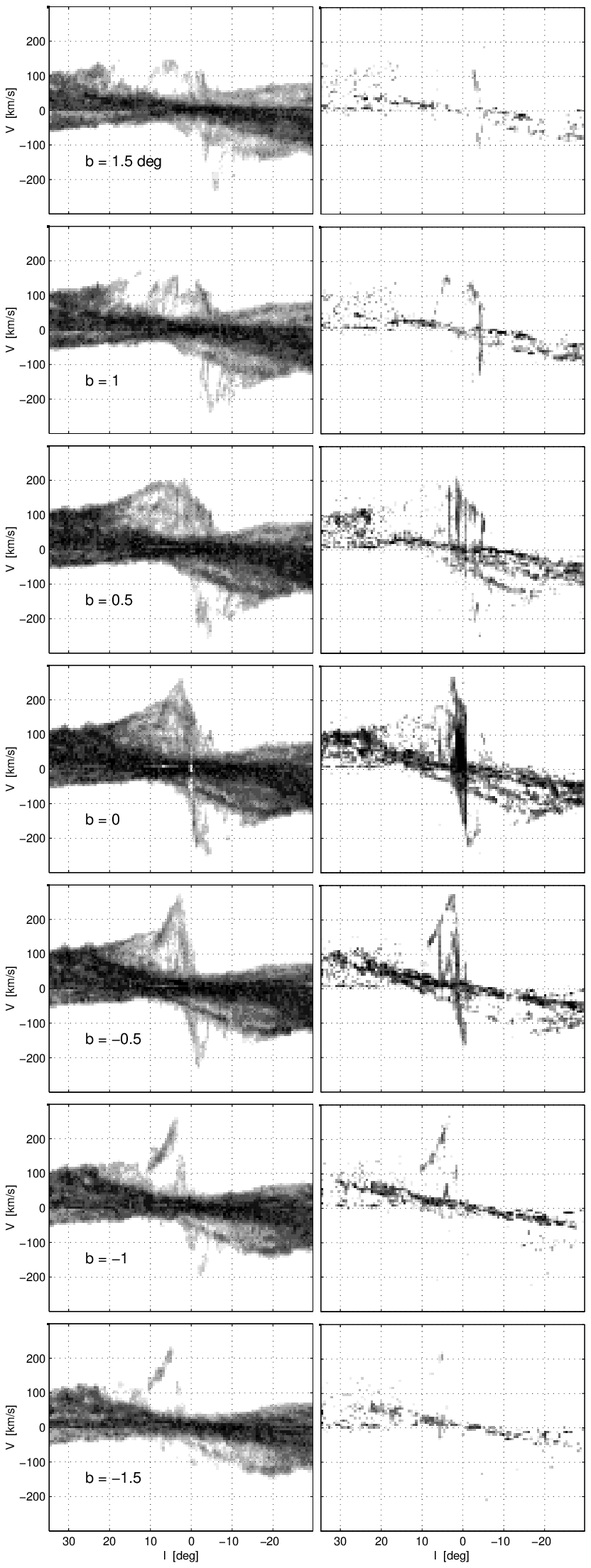,width=8.8cm}}
\end{figure}
All features listed here appear both in the CO and HI data with
a~rigorous coincidence in position, although the central molecular
zone is much weaker in HI. It is not yet clear whether these features
are transient or represent a permanent gas flow on closed orbits.
\par Bania~(1977) has also isolated two particular ``clumps'' in the
$^{12}$CO data (see Fig.~\ref{lv}), corresponding to massive molecular
cloud complexes.
\begin{itemize}
\item {\it Clump~1} is located at the negative longitude end of the
\mbox{135-km\,s$^{-1}$} arm, at $\ell\approx -5\degr$ and
$b\approx 0.4\degr$, and represents the most extreme case of
non-circular motion in the Milky Way's dense gas kinematics. It is
located at slightly lower latitude than the nearby emission from the
135-km\,s$^{-1}$ arm. The clump~1 complex has been resolved into three
distinct sub-complexes each of about $2.5\times 10^5$~$M_{\odot}$
H$_2$ mass. More details on its physical properties are presented in
Bania et al.~(1986).
\item {\it Clump~2} is confined near $\ell =3\degr$ and $b=0.2\degr$
and spans an extraordinary large velocity range of over
$150$~km\,s$^{-1}$. It has been resolved into 16 emitting cores of
$\sim 5\times 10^5$~$M_{\odot}$ by Stark \& Bania~(1986), who also
interpret this clump as a dustlane or an inner spiral arm in a barred
potential formed by molecular clouds distributed along the line of
sight.
\end{itemize}
\par The CO is usually considered as a tracer of H$_2$, although it is
now well established that the density ratio between these two
molecules is not homogeneous in galaxies. Beside the $2.6$~mm
$^{12}$CO $J=1\rightarrow 0$ emission line, other more transparent
molecular tracer like $^{12}$CO $J=2\rightarrow 1$ (Oka et al. 1998a),
$^{13}$CO and CS (Bally et al. 1987), as well as HCN (Jackson et
al. 1996; Lee 1996), have also been used to map dense gas like in the
Galactic centre region.
\par The interpretation of features in the $\ell-V$ diagrams is
subject to many difficulties. First, since only the radial component
of the velocity is measured, the in-plane velocity field can be
recovered directly only under very restrictive symmetry assumptions
like axisymmetry. Second, some features may not trace real spiral arms
but be artifact of velocity crowding along the line of sight (Burton
1973; Mulder \& Liem 1986). Third, features elongated in the velocity
direction can be understood either by real structures elongated along
the line of sight or by spatially localised emission with a very large
velocity spread, as would result for example in a supernova explosion
or in violent shocks. Finally, some regions in the $\ell-V$ space
overlay the emission from several distinct sources, in particular the
low velocity emission from the Galactic centre region is partly hidden
by the emission from the surrounding disc spiral arms. This problem can
however be addressed considering the latitudinal gas distribution or
resorting to very dense gas tracers. Optically thick regions are also
strongly affected by absorption.

\section{Numerical methods}
%
The initial models are evolved using a composite $N$-body and hydro
code developed by the Geneva Observatory galactic dynamic group,
originally provided by D. Friedli but considerably modified in the
present application. At each time step, the gravitational forces on
all particles (gas included) are derived by the particle-double mesh
method detailed in Sect.~\ref{pm2}, the pressure and viscous forces on
the gas particles by the Lagrangian smoothed particle hydrodynamics
(SPH) technique described in Sect.~\ref{SPH}, and finally the phase
space coordinates of each particle are updated by integration of the
equations of motion according to the algorithm presented in
Sect.~\ref{int}.
\begin{figure}[t!]
\centerline{\psfig{figure=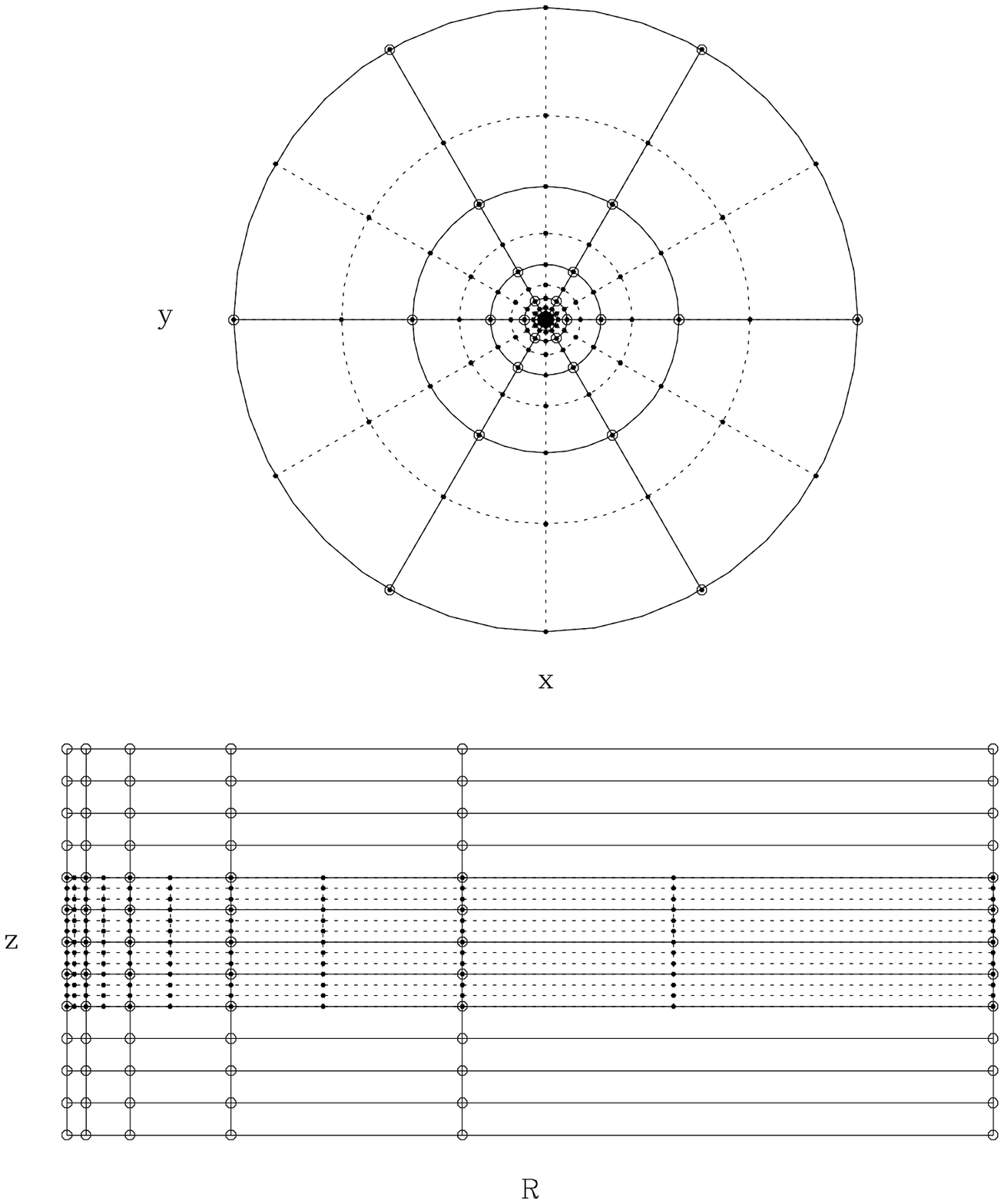,width=8.8cm}}
\caption{Example of a double grid with $N_R=10$, $N_{\phi}=12$,
         $N_z=13$ and $M_z=3$. The solid points and dotted lines are
         the points and meshes of grid~A, and the circles and solid
         lines those of grid~B.}
\label{double1}
\end{figure}

\subsection{Gravitation with PM$^2$}
\label{pm2}
To compute the gravitational forces on the particles, we have resorted
as in paper I to the particle-mesh (PM) technique with
polar-cylindrical grid geometry and variable homogeneous ellipsoidal
kernel for the softening of the short range forces (Pfenniger \&
Friedli 1993). However, instead of using a single grid with constant
vertical resolution, inappropriate to model efficiently the high
density contrast between a thin disc and an extended dark halo, we
have introduced a double embedded grid structure. The higher
resolution grid~(A) has $N_R\times N_{\phi}\times N_z$ cells, with a
maximum radial extent $R_{\rm max}$ and a vertical spacing $H_z$, i.e.
$z^{\rm A}_{\rm max}=(N_z-1)/2\cdot H_z$. The lower resolution
grid~(B) has half the number of cells in each dimension of the plane,
skipping every second $R$- and $\phi$-grid point of the finer grid,
and the same number of cells vertically but with a spacing of
$M_z\cdot H_z$, i.e. $z^{\rm B}_{\rm max}=M_z\cdot
z^{\rm A}_{\rm max}$. Hence grid~B has the same radial extent than
grid~A, but is larger in the vertical dimension. Both grids have
constant vertical spacing, thus allowing to keep the advantage of the
Fast Fourier Transform algorithm in this dimension. Here $N_z$
designates the number of active cells in~$z$, corresponding to half
the value required by the doubling-up method (Hockney \& Eastwood
1981). The vertical grid parameters are chosen to satisfy:
\begin{equation}
N_z=1+2k\cdot M_z,\hspace*{.8cm} k\;\:{\rm integer}\geq 2,
\end{equation}
ensuring that (i) there are grid points of each grid in the plane
$z=0$, (ii) the horizontal planes $z=\pm z^{\rm A}_{\rm max}$
delimiting vertically grid~A also contain points of grid~B, and
(iii)~grid~B has at least four cells within grid~A in the $z$
dimension. Figure~\ref{double1} gives an example of double grid
fulfilling all these conditions.
\begin{figure}[t!]
\centerline{\psfig{figure=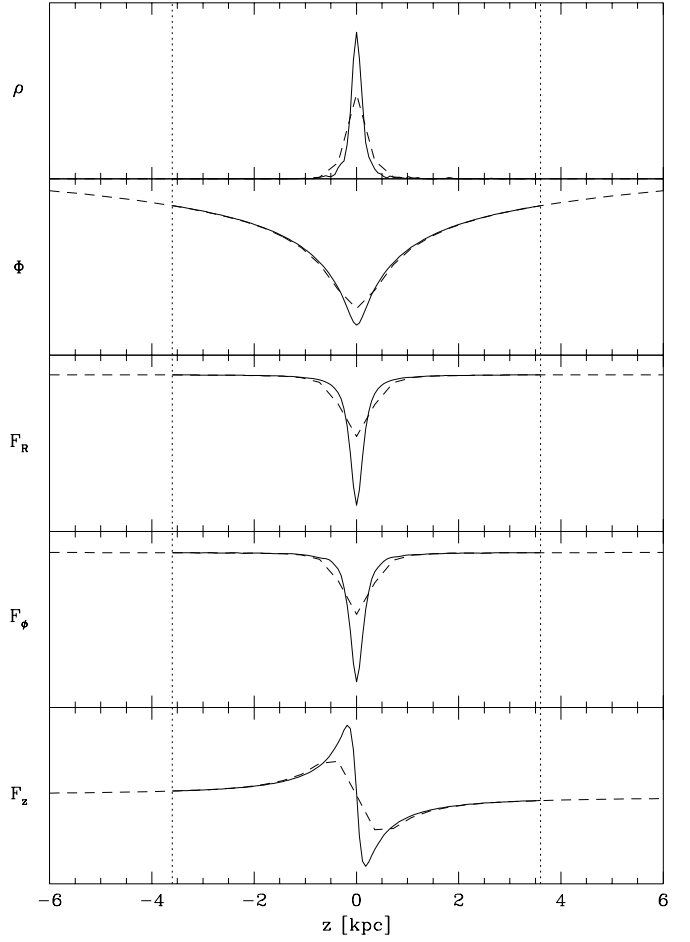,width=8.8cm}}
\caption{Typical $z$-behaviour of model dynamical properties
         calculated with the high resolution double grid, i.e.
         $N_R=62$, $N_{\phi}=64$, $N_z=121$, $M_z=6$,
         $R_{\rm max}=50$~kpc and $H_z=60$~pc (the mass distribution
         and the grid refer to model l10t2000). From top to bottom and
         in linear scales: mass density ($\rho$), potential ($\Phi$)
         and the three force components ($F_R$, $F_{\phi}$ and $F_z$).
         The full lines are the results from grid~A (based on
         $\Phi_{\rm A}+\Phi_{\rm C}$), and the dashed lines those from
         grid~B (based on $\Phi_{\rm B}$). The vertical dotted lines
         indicate the limits $z=\pm z_{\rm max}^A$ of grid~A. The
         forces considerably differ from one grid to the other near
         the plane $z=0$, but almost confound in the transition zone.}
\label{double}
\end{figure}
\par In this particle-double mesh (or PM$^2$) method, the mass
assignment to the grid cells is repeated twice (once for each grid),
and the total potential involves the evaluation of three
sub-potentials: $\Phi_{\rm A}$ and $\Phi_{\rm B}$ generated by the
mass inside grids~A and B respectively, as well as $\Phi_{\rm C}$
generated by the mass within grid~B but outside grid~A. The potential
$\Phi_{\rm A}$ is computed on grid~A, at high resolution, and the
other potentials on grid~B. The total potential then amounts to
$\Phi_{\rm A}+\Phi_{\rm C}$ in the region of grid~A, which requires
an interpolation of $\Phi_{\rm C}$ to grid~A, and simply to
$\Phi_{\rm B}$ outside this region. The increase of CPU time owing to
the triple potential evaluation is amply compensated by the reduction
by a factor $\sim M_z$ of the effective number of cells relative to a
single grid with the same resolution as grid~A and the same size as
grid~B. The transition planes between the two grids are associated
with numerical discontinuities in the force components, although quite
small in practice (see Fig.~\ref{double}). To smooth them out, the
forces in the first and last cells of grid~B within grid~A are taken
as a linear combination so that the forces at
$z=\pm z^{\rm A}_{\rm max}$ coincide with the forces of grid~B and
those at $z=\pm(z^{\rm A}_{\rm max}-M_zH_z)$ with the forces of
grid~A.
\par The grid parameters adopted in our simulations are given in
Table~\ref{simul} and the resulting gravitational resolution achieved
with the high resolution double grid is depicted in Fig.~\ref{resol}.

\subsection{Gas hydrodynamics with SPH}
\label{SPH}
The pressure and viscous forces acting on the gas particles are
derived by three-dimensional SPH. We will not repeat here the
principle of this very popular Lagrangian method for solving Euler's
equation of motion (see Benz 1990; Monaghan 1992 and Steinmetz 1996
for reviews), but just give the necessary specifications about this
part of our code.
\par The relevant form of the Euler equation writes:
\begin{equation}
\frac{d\vec{v}}{dt}\equiv \frac{\partial{\vec{v}}}{\partial t}
+(\vec{v}\cdot\vec{\nabla})\vec{v}=
-\frac{\vec{\nabla}P}{\rho}-\Pi-\vec{\nabla}\Phi,
\label{euler}
\end{equation}
where $\vec{v}(\vec{r})$ is the velocity field, $\rho(\vec{r})$ the
spatial mass density and $P(\vec{r})$ the pressure of the gas. $\Pi$
is the viscosity term and $-\vec{\nabla}\Phi$ the gravitational force
as obtained by the PM$^2$ method.
\par The internal properties of a classical fluid are fully described
by three macroscopic independent variables of state, like for example
the density and the pressure appearing explicitly in Eq.~(\ref{euler}),
and the internal energy per unit mass~$u(\vec{r})$. The density is
evaluated directly at each particle position $\vec{r}_i$ via:
\begin{equation}
\rho_i\equiv \rho(\vec{r}_i)
      =\sum_{j=1}^{N_{\rm g}}m_j W(\vec{r}_i-\vec{r}_j,h),
\label{rho}
\end{equation}
where $N_{\rm g}$ is the number of gas particles, $m_j$ the individual
mass of the particles, $W(\vec{r},h)$ the kernel function and $h$ the
smoothing length. In SPH, this equation is equivalent to the
continuity equation expressing the conservation of total mass.
\par A first relation between the variables of state is given by the
equation of state assuming a perfect gas, i.e.
$P=(\gamma-1)\rho \cdot u$, where $\gamma$ is the adiabatic index of
the gas, which from statistical mechanics is given by $\gamma=1+2/f$,
with $f$ being the number of freedom degrees of the gas constituents
($f=3$ for atomic hydrogen and $f=5$ for molecular hydrogen). A second
relation comes from the energy conservation equation, which in its
general form involves cooling and heating functions. Since these
functions are poorly constrained by theory, we have simply resorted to
the isothermal assumption, i.e. the internal energy of the gas is at
any time and everywhere the same. This assumption, partly justified by
the homogeneous velocity dispersion of the warm gas component in
external galaxies (e.g. van der Kruit \& Shostak 1984), means that
cooling and heating exactly cancel each other and that the energy
released by the shocks is instantaneously radiated away. Hence,
introducing the (constant) sound speed
$c_{\rm s}=[\gamma(\gamma-1)\cdot u]^{1/2}$ to substitute the internal
energy, the equation of state for particle~$i$ reduces to:
\begin{equation}
P_i=\frac{c_{\rm s}^2}{\gamma}\cdot \rho_i,
\end{equation}
which is degenerated in sound speed and adiabatic index, depending
only on the ratio $c_{\rm s}^2/\gamma$. With this last equation,
Benz's~(1990) SPH version of the pressure term in Euler's equation
simply becomes:
\begin{equation}
\left(\frac{\vec{\nabla}P}{\rho}\right)_i\approx -\frac{c_{\rm s}^2}{\gamma}
\cdot \sum_{j=1}^{N_{\rm g}}m_j\left(\frac{1}{\rho_i}\!+\!\frac{1}{\rho_j}
\right)\vec{\nabla}_i W(\vec{r}_i-\vec{r}_j,h).
\label{P}
\end{equation}
\par The empirical artificial viscosity is calculated exactly as in
Benz~(1990). It consists of a superposition of the ``bulk'' and the
von Neumann-Richtmyer viscosities, with the standard parameters set to
$\alpha=1.0$ and $\beta=2.5$, and takes into account Balsara's~(1995)
correction to avoid energy dissipation in pure shearing flows.
\begin{figure}[t!]
\centerline{\psfig{figure=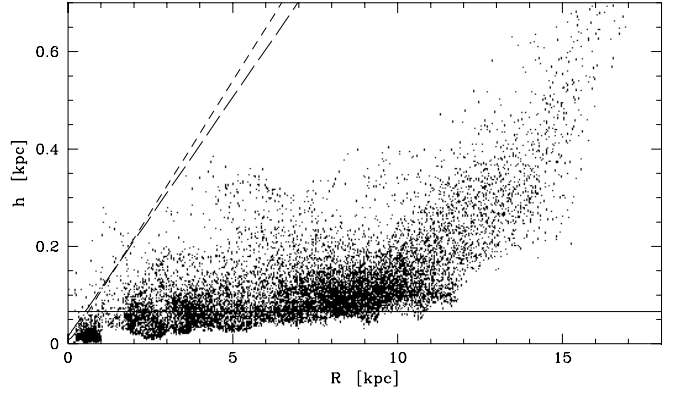,width=8.8cm}}
\caption{Various spatial resolutions of the high resolution
         simulations presented in this paper, as a function of
         galactocentric radius. The solid, short dashed and long
         dashed lines respectively represent the vertical, azimuthal
         and radial semi-axes of the gravitational homogeneous
         ellipsoidal kernel associated with grid~A, and the points are
         the smoothing lengths of a random selection of $10$\% SPH
         particles in model l10t2000. The gas smoothing length is
         close to the vertical gravitational resolution.}
\label{resol}
\end{figure}
\par We have used the spherical spline kernel because of its finite
spatial extension:
\begin{equation}
W(\vec{r},h)=\frac{1}{\pi h^3}
\left\{\begin{array}{ll}
       1-\frac{3}{2}w^2+\frac{3}{4}w^3 & {\rm if}\hspace{.3cm} 0\leq w\leq 1,\\
       \frac{1}{4}(2-w)^2              & {\rm if}\hspace{.3cm} 1\leq w\leq 2,\\
       0                               & {\rm if}\hspace{.3cm} w\geq 2,
       \end{array}\right.
\end{equation}
where $w\equiv |\vec{r}|/h$. To increase the hydro resolution in high
density regions like shocks, the particles are assigned individual
smoothing lengths~$h_i$ such that their number of neighbouring
particles~$N_i$ always remains close to a fixed number $N_{\circ}$.
Two particles $i$ and $j$ are defined as mutual neighbours if
$i\neq j$ and $|\vec{r}_i-\vec{r}_j|<2h_{ij}$, where
$h_{ij}=(h_i+h_j)/2$ is the symmetrised smoothing length which should
enter the kernel evaluations in Eqs.~(\ref{rho}) and~(\ref{P}) to
ensure momentum conservation.
\begin{figure}[t!]
\centerline{\psfig{figure=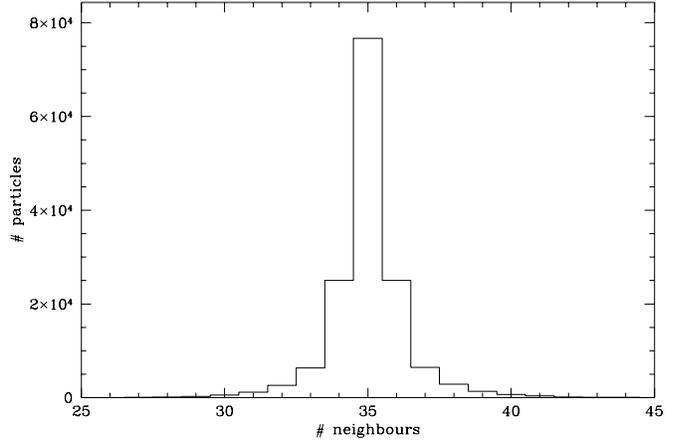,width=8.8cm}}
\caption{Typical distribution of the number of neighbouring SPH
         particles, obtained by adjusting at each time step the
         smoothing lengths according to Eqs.~(\ref{hdot})
         and~(\ref{new}). The data refer to model l10t2000.}
\label{voisin}
\end{figure}
\par At each time step, the smoothing lengths are updated considering
the general three-dimensional scaling law:
\begin{equation}
\frac{h_i}{h_{\circ}}=\left(\frac{N_i+1}{N_{\circ}+1}\frac{\rho_{\circ}}
                                 {\rho_i}\right)^{1/3},
\end{equation}
where $h_{\circ}$ and $\rho_{\circ}$ are constants, and $+1$ is added
to account for particle $i$. Following Benz~(1990), differentiating
this formula and substituting the continuity equation yields:
\begin{equation}
\dot{h_i}\equiv \frac{dh_i}{dt}
               =\frac{1}{3}h_i\left(\frac{1}{N_i+1}\frac{dN_i}{dt}
                +[\vec{\nabla}\cdot \vec{v}]_i \right).
\label{hdot}
\end{equation}
Benz did only consider the variation of $h_i$ ensuring the constancy
of the neighbour numbers, and hence did not include the term in $N_i$.
However, in practice, numerical fluctuations of the $N_i$'s can carry
these number outside a reasonable range around~$N_{\circ}$. To
overcome this problem, we simply damp the systematic departure of the
$N_i$'s from $N_{\circ}$ setting:
\begin{equation}
\frac{dN_i}{dt}=\frac{N_i-N_{\circ}}{\eta \Delta t}
\label{new}
\end{equation}
in Eq.~(\ref{hdot}) and integrate the resulting equation along with the
equations of motion (see Sect.~\ref{int}). The parameter $\eta$
controls the damping rate per time step ($\Delta t$) and should be
significantly greater than $1$ to avoid abrupt discontinuities in the
non-gravitational forces. In all simulations we have adopted
$N_{\circ}=35$ and $\eta=5$. Figure~\ref{resol} shows the typical SPH
resolution achieved in evolved models and Fig.~\ref{voisin} an example
of the neighbour number distribution. The standard deviation of this
distribution usually amounts to only about $1.5$ neighbours. The
velocity divergence in Eq.~(\ref{hdot}), which is also needed for the
viscous forces, is estimated as in Benz~(1990).
\par The initial smoothing lengths of the gas particles, confined near
the plane $z=0$, are derived from the two-dimensional scaling
relation:
\begin{equation}
h_i(t=t_{\circ})=\sqrt{\frac{N_{\circ}m_i}{4\pi \Sigma_{\rm g}(R_i)}},
\end{equation}
where $\Sigma_{\rm g}(R)$ is the surface mass density profile of the
gas component and $t_{\circ}$ its switch-on time (see
Sect.~\ref{time}). This method unfortunately leads to a large spread
in the initial neighbour numbers. To adjust the $h_i$'s without
evolving too much the system, the simulations with live gas are
therefore started with a time step much smaller than in the subsequent
stabilised regime, i.e. $\Delta t=0.001$~Myr for the large simulations
(l-series) and $\Delta t=0.01$~Myr for the smaller ones (s-series).

\subsection{Integrator}
\label{int}
To integrate the equations of motion, we have applied a
semi-adaptative second order accurate algorithm where the (Cartesian)
phase space coordinates, contrary to the standard leap-frog scheme,
are evaluated at synchronous times. At each time step the positions
$\vec{r}_i$ and velocities $\vec{v}_i$ of the particles are modified
according to (e.g. Hut et al. 1995):
\begin{eqnarray}
\vec{r}^{\, n+1}_i & = & \vec{r}^{\, n}_i+\vec{v}^{\: n}_i\Delta t^n
                                 +\frac{1}{2}\vec{a}^{\, n}_i(\Delta t^n)^2, \\
\vec{v}^{\: n+1}_i & = & \vec{v}^{\: n}_i+\frac{1}{2}(\vec{a}^{\, n}_i
                                 +\vec{a}^{\, n+1}_i)\Delta t^n, \\
h^{n+1}_i & = & h^n_i+\frac{1}{2}(\dot{h}^n_i+\dot{h}^{n+1}_i)\Delta t^n,
\end{eqnarray}
where the $\vec{a}_i$'s are the accelerations and $\Delta t$ is the
time step, and where the last equation is used to update the smoothing
lengths of the SPH particles. The indices $n$ and $n+1$ refer to the
values at time $t^n$ and $t^{n+1}=t^n+\Delta t^n$ respectively.
\par
\begin{figure}[t!]
\centerline{\psfig{figure=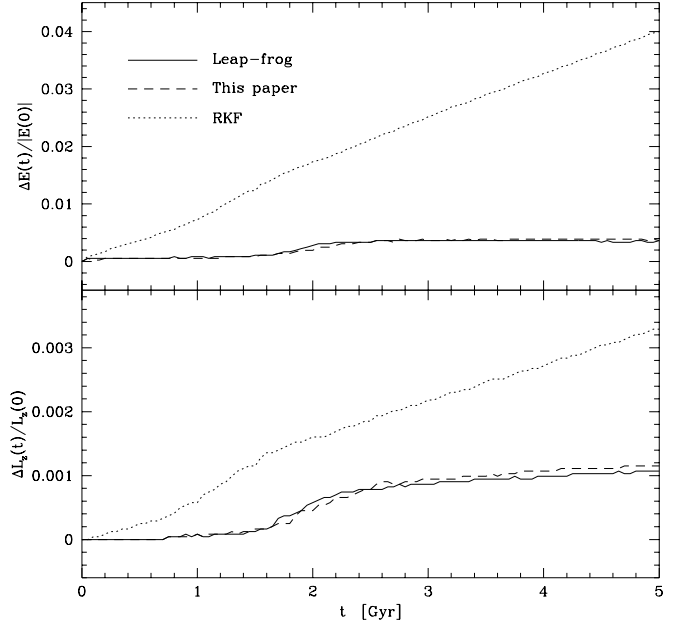,width=8.8cm}}
\caption{Variation of the total energy (top) and of the total angular
         momentum about the rotation axis (bottom) for simulations
         started from the same initial conditions but evolved using
         three different second order integrators with the same
         constant time step; $\Delta E(t)\equiv E(t)-E(0)$ and
         $\Delta L_z(t)\equiv L_z(t)-L_z(0)$. The steeper part of the
         curves are associated with the epoch of bar formation.}
\label{cons}
\end{figure}

 The first equation handling the positions is not exactly
reversible in time. However, we have repeated simulation m00 of
paper~I replacing successively the leap-frog integrator by this
algorithm and a second order Runge-Kutta-Fehlberg algorithm (RKF;
Fehlberg 1968) as in Friedli's original code, and found that, for the
same constant time step $\Delta t=0.1$~Myr, the conservation of the
total energy (and of the total angular momentum) is similar for the
present algorithm and for the leap-frog, but much worse for the RKF
($\Delta E/|E|$ larger by a factor $\sim 20$, see Fig.~\ref{cons}).
Also, the bar begins to form much sooner with the RKF integrator, at
$t\sim 800$~Myr instead of $\sim 1600$~Myr with the other two
integrators, which is a sign of higher numerical noise amplification.
\par For the gaseous particles, the pressure and viscous forces 
contributing to $\vec{a}^{\, n+1}_i$ both involve velocities which are
not known a priori, and similarly $\dot{h}^{n+1}_i$ depends on
$h^{n+1}_i$ (through Eq.~(\ref{hdot})). Therefore the
non-gravitational part of $\vec{a}^{\, n+1}_i$ and $\dot{h}^{n+1}_i$
are calculated with the first order predictors:
\begin{eqnarray}
\vec{v}^{\: {\rm p}}_i & = & \vec{v}^{\: n}_i+\vec{a}^{\, n}_i\Delta t^n, \\
h^{\rm p}_i & = & h^n_i+\dot{h}^n_i\Delta t^n,
\end{eqnarray}
which fortunately preserve the second order accuracy of
$\vec{r}^{\, n+1}_i$, $\vec{v}^{\: n+1}_i$ and $h^{n+1}_i$.
\par To temporally resolve the high density shocks in the gas
component, we use an adaptative time step procedure inspired from
Friedli~(1992). After step $n$, the next time step $\Delta t^{n+1}$ is
estimated from the maximum relative contribution per unit time of the
second order terms to the integrated quantities:
\begin{equation}
\epsilon_{\rm max}^{n+1}=\begin{array}[t]{c}{\rm Max} \vspace{-.2cm}\\
\hbox{\scriptsize $i$}\end{array}\!\!\left\{\epsilon(\vec{r}^{\, n+1}_i),\:
\epsilon(\vec{v}^{\: n+1}_i),\:\epsilon(h^{n+1}_i)\right\}, \label{eps}
\end{equation}
where
\begin{eqnarray}
\epsilon(\vec{r}^{\, n+1}_i) & = & \frac{1}{2}\frac{|\vec{v}^{\: n+1}_i
-\vec{v}^{\: n}_i|}{\sqrt{|\vec{r}^{\, n+1}_i|^2+r_{\rm lim}^2}},\\
\epsilon(\vec{v}^{\: n+1}_i) & = & \frac{1}{2}\frac{|\vec{a}^{\, n+1}_i
-\vec{a}^{\, n}_i|}{\sqrt{|\vec{v}^{\: n+1}_i|^2+v_{\rm lim}^2}},\\
\epsilon(h^{n+1}_i) & = & \frac{1}{2}\frac{|\dot{h}^{n+1}_i-\dot{h}^n_i|}
{\sqrt{(h^{n+1}_i)^2+h_{\rm lim}^2}},
\end{eqnarray}
and where $r_{\rm lim}=h_{\rm lim}$ and $v_{\rm lim}=h_{\rm lim}/
\Delta t^n$ are destinated to prevent divergences and a drastic drop
of the time step. The value of $h_{\rm lim}$ is fixed to $0.01$~pc in
the large simulations and to $0.02$~pc in the smaller ones,
corresponding approximatively to the smallest dimension of the PM$^2$
cells. Ideally, to guarantee a relative uncertainty of the same order
or smaller than a given tolerance $E_{\rm tol}$, it suffice to take
$\Delta t^{n+1}=E_{\rm tol}/\epsilon_{\rm max}^{n+1}$. However, a more
advisable prescription is:
\begin{equation}
\Delta t^{n+1}={\rm Max}\left\{{\rm Min}\left[\alpha^{n+1}\cdot \Delta t^n,\:
               \Delta t_{\circ}\right],\:\Delta t_{\rm min}\right\},
\end{equation}
with
\begin{eqnarray}
\alpha^{n+1} & = & {\rm Min}\left\{\sqrt{\beta^{n+1}},\:
                   \alpha_{\rm max}\right\}, \\
\beta^{n+1}  & = & \frac{E_{\rm tol}}{\epsilon_{\rm max}^{n+1}\Delta t^n},
\end{eqnarray}
and $\alpha_{\rm max}=2$. The square root over $\beta^{n+1}$ softens
the variations of the time step and $\alpha_{\rm max}$ puts an upper
limit on its growth. Moreover the time step is maintained in the range
$\Delta t_{\rm min}\leq \Delta t\leq \Delta t_{\circ}$, with
$\Delta t_{\rm min}=10^{-3}$~Myr and $\Delta t_{\circ}=0.1$~Myr. The
upper boundary is a fraction of the time needed by the stars to cross
the central high resolution PM$^2$ cell in the steep nuclear
potential. This time-stepping criterion is applied only to the gas,
i.e. the loop on $i$ in Eq.~(\ref{eps}) is restricted to the gas
particles, and $\epsilon_{\rm max}^{n+1}$ is in general controlled by
the smoothing lengths. 
\par If $\beta<2/3$ and $\Delta t^n>\Delta t_{\rm min}$, the tolerance
is considered as not respected in the last integration step. In this
case, the old values of the integrated quantities are restored and 
integrated again with a time step
$\Delta t^n={\rm Max}\{\sqrt{\beta^{n+1}}\cdot
\Delta t^n,\:\Delta t_{\rm min}\}$. This rejection procedure
unfortunately requires to store the old positions of all particles.
\par In the simulations with live gas, the tolerance has been set to
$E_{\rm tol}=10^{-1}$, and in the simulations with fixed gas
component, a constant time step $\Delta t=\Delta t_{\circ}$ has been
applied.

\section{Initial conditions}
%
The radial distribution of molecular gas in the Milky~Way, as traced
by CO, is essentially confined inside the solar circle, with a marked
hole between $1$ and $3$~kpc and a strong central concentration
associated to the nuclear ring/disc, whereas the HI presents a rather
constant surface density extending far beyond $R_{\circ}$ and an
abrupt decline towards the centre below $3$~kpc. The total gas mass
within $R_{\circ}$ is estimated to $4.2\times 10^9$~$M_{\odot}$,
comprising $2.3\times 10^9$~$M_{\odot}$ of H$_2$ (with a large
uncertainty owing to the poorly known CO to H$_2$ conversion factor)
and $0.7\times 10^9$~$M_{\odot}$ of HI (Scoville 1992), plus $28$\% of
helium and metals. For the gaseous disc, we have used the following
initial mass distribution:
\begin{equation}
\rho_{\rm g}(R,z)=\frac{M_{\rm g}}{(2\pi)^{3/2}\sigma^2pR}
                  \exp{\left[-\frac{1}{2}\left(\frac{R^2}{\sigma^2}
                  +\frac{z^2}{p^2R^2}\right)\right]},
\end{equation}
with $\sigma=5$~kpc, $p=0.017$ and a total mass
$M_{\rm g}=5\times 10^9$~$M_{\odot}$, resulting in
$3.6\times 10^9$~$M_{\odot}$ within the solar circle. Both radial and
vertical profiles are Gaussian, and the disc is linearly flaring with
radius, as observed in HI from a few~kpc of the centre out to at least
$2R_{\circ}$ (Merrifield 1992), achieving a thickness of $136$~pc at
$R=R_{\circ}$. The fast radial decline is aimed to spare SPH particles
in the outer regions and hence increase the spatial resolution near
the centre at fixed number of particles. The observed gas deficit
between $1$ and $3$~kpc needs not to be reproduced since shocks will
naturally deplete this region. The gas particles initially have pure
circular velocities derived from the axisymmetric part of the total
potential at $z=0$.
\par The stellar and dark mass is divided into the same three
components as in paper I: a stellar nucleus-spheroid (NS), a stellar
disc and a dark halo (DH), with their initial axisymmetric mass
distribution described by the same analytical formulae, except for the
disc. In paper I we indeed noted that the adopted double exponential
discs evolve into discs with too less mass outside the bar region
according to the COBE/DIRBE near-IR and bulge microlensing data, and
possibly an excess of mass in the central region when compared to the
HI terminal velocity constraints. Instead of increasing the disc scale
length, we choose here to soften the initial central mass density
taking:
\begin{eqnarray}
\rho_{\rm d}(R,z) & = & g(z/h_z)\frac{M_{\rm d}}{2\pi h_R^2 h_z[\exp{(-1)}
                                                  +\exp{(-1/2)}]}\nonumber \\
& & \hspace{.3cm}\cdot
\left\{\begin{array}{ll}
       \exp{[-\frac{1}{2}(1+\frac{R^2}{h_R^2})]} & \hspace{.5cm}{\rm if}
                                                   \hspace{.3cm}R\leq h_R, \\
       \exp{[-\frac{R}{h_R}]} & \hspace{.5cm}{\rm if}\hspace{.3cm} R\geq h_R,
       \end{array} \right.
\end{eqnarray}
where $M_{\rm d}$ is the total disc mass, $h_R$ the scale length,
$h_z$ the scale height and $g(\zeta)$ the normalised vertical profile.
The surface density remains exponential in the external disc, but
continuously and differentiably joins an inner Gaussian distribution
at $R=h_R$, with a central value reduced by $40$\% relative to the
purely exponential case. To compensate for the enhanced spatial
density near the plane $z=0$ caused by the additional gas component,
we also replace the exponential vertical profile of paper~I by
van~der~Kruit's~(1988) profile:
\begin{equation}
g(\zeta)\propto {\rm sech}^{2/n}\!\left[\frac{n}{2}\zeta\right],
\end{equation}
with $n=7$, between the exponential ($n=\infty$) and isothermal
($n=1$) cases.
\par The choice of the parameters are based on the simulations
performed in paper I, giving a strong weight to the $x_1$ orbits
versus HI terminal velocities test. The best models regarding this
test are m06t4600 and m04t3000, whose initial axisymmetric models
share the following interdependent properties: (i) a ratio of disc
over NS mass of $0.8$ within the spheroidal volume $s<3$~kpc (where
$s^2=R^2+z^2/e^2$ and $e=0.5$), (ii)~a total NS+disc mass of
$2.4\times 10^{10}$~$M_{\odot}$ in the same volume and (iii) a
circular velocity of $190$~km\,s$^{-1}$ just after the very steep
central rise of the rotation curve, taking into account the velocity
scales adjusted to the observed stellar velocity dispersion in Baade's
window (see Table 3 of paper I). The DH parameters, $a$ and $h_z$ are
fixed as in the reference simulation m00, and the other parameters are
adjusted to the former constraints and to a disc surface density on
the solar circle of $60$~$M_{\odot}$\,pc$^{-2}$, yielding
$M_{\rm NS}= 2.48\times 10^{10}$~$M_{\odot}$,
$M_{\rm d}=4.6\times 10^{10}$~$M_{\odot}$ and $h_R=3.2$~kpc. The mass
density is softly truncated as in paper I, at radius
$R_{\rm c}=38$~kpc and over a width $\delta=5$~kpc, and the initial
kinematics rests on the same relations between the velocity moments as
in simulations m00-m10, except that the asymptotic velocity anisotropy
$\beta_{\infty}=0.68$ and the transition radius $r_{\circ}=5.3$~kpc
for the NS component.
\begin{table*}[t!]
\centering
\caption{List and characteristics of the simulations presented in this
         paper. $N_R$, $N_{\phi}$, $N_z$ and $H_z$ are the grid
         parameters as defined in Sect.~\ref{pm2}. Both low and high
         resolution double grids have $M_z=6$, $R_{\rm max}=50$~kpc,
         $z^{\rm A}_{\rm max}=3.6$~kpc and
         $z^{\rm B}_{\rm max}=21.6$~kpc. \# is the number of
         particles, $c_{\rm s}$ the sound speed of the gas,
         $t_{\circ}$ the gas switch-on time and $t_{\rm end}$ the time
         to which the simulation is performed. The points refer to the
         former lines.}
\begin{tabular}{cccccccccccc} \hline
Simulation & $N_R$ & $N_{\phi}$ & $N_z$ & $H_z$ [pc] & \# NS & \# disc
& \# DH & \# gas &
$t_{\circ}$ [Myr] & $t_{\rm end}$ [Myr] & $c_{\rm s}$ [km\,s$^{-1}$] \\
                                                                       \hline
sxx & 32 & 32 & ~73 & 100 & ~83\,444 & ~\,178\,022 & ~\,225\,284 &        0 &  --  & 5000 & -- \\
s10 &  . &  . &  .  &  .  &    .     &      .      &      .      & ~20\,000 & 3200 &   .  & 10 \vspace{.3cm}\\
lxx & 62 & 64 & 121 & ~60 & 625\,828 & 1\,335\,167 & 1\,689\,629 &        0 &  --  &   .  & -- \\
l10 &  . &  . &  .  &  .  &    .     &      .      &      .      & 150\,000 & 1800 & 2275 & 10 \\
l20 &  . &  . &  .  &  .  &    .     &      .      &      .      &    .     &  .   &   .  & 20 \\
\hspace{.1cm}l10' & . & . & .  & . & . &    .      &      .      &    .     & 2400 & 2775 & 10\\ \hline
\label{simul}
\end{tabular}
\end{table*}

\section{Time evolution}
\label{time}
Table~\ref{simul} gives an overview of the simulations described in
this paper. Two series of simulations have been realised: the first
one with moderate number of particles and spatial resolution (series
``s'', small simulations), and the other one pushing these quantities
such as to exploit about half of the memory resources of the most
powerful computers available at Geneva University (series ``l'', large
simulations). In each series, we first integrated the initial
axisymmetric model for $5$ Gyr keeping the gaseous component fixed
(simulations ``xx''), and then we relaxed the gas particles at
different intermediate times $t_{\circ}$. The gas is not evolved from
the beginning because the non-inclusion of star formation produces an
excessive accumulation of gas in the central region due to the torques
of the non-axisymmetric gravitational potential, raising the rotation
curve and leading to a premature destruction of the bar (Friedli \&
Benz 1993) or even preventing its formation. Instead of introducing an
artificial gas recycling procedure, the time consuming gas-live
simulations of the l-series were integrated only over a few $100$~Myr.
To damp the initial disequilibrium of the gas owing to its circular
kinematics in the already barred potential, the non-axisymmetric part
of the potential is progressively and linearly brought to its nominal
value in $75$~Myr, i.e. roughly half a rotation period of the bar.
\par In the l-series, the gas has been released at two different times
after the formation of the bar, at $t_{\circ}=1.8$ and $2.4$~Gyr. In
the first case, two values of the sound speed have been explored,
$c_{\rm s}=10$ and $20$~km\,s$^{-1}$ (simulations l10 and l20
respectively), and in the second case only $c_{\rm s}=10$~km\,s$^{-1}$
has been retained (simulation l10'). In the s-series, many runs with
live gas have been performed, releasing the gas every $400$~Myr and
each time with two different sound speeds, but only the one mentioned
in Table~\ref{simul} will be discussed here. The adiabatic index of
the gas is set to that of neutral hydrogen, i.e. $\gamma=5/3$. Taking
the vertical gravitational resolution $H_z$ as a lower physical limit
for the SPH smoothing length, the maximum gaseous density that can be
reasonably modelised in the l-simulations~is
$\sim W(0,H_z)\cdot N_{\circ}/N_{\rm g}\cdot M_{\rm g}\approx
2$~$M_{\odot}$\,pc$^{-3}$ (see the previous sections for the meaning
of the symbols), which is at least one order of magnitude below the
density of the Galactic molecular gas, but sufficient to describe the
warmer neutral phase with a typical sound speed of $10$~km\,s$^{-1}$.
However, Cowie~(1980) has argued that a system of molecular clouds may
be treated as a classical fluid with a sound speed equal to the mass
averaged cloud velocity dispersion, which also amounts to
$10$--$20$~km\,s$^{-1}$ in the Milky~Way. Hence our simulations are
expected to display properties of both the HI and H$_2$ medium.
\par The SPH particles all have the same time-independent mass. The
number of particles in each luminous component is always such that the
mass per particle is the same as for the gas component to minimise
relaxation effects, and the number of DH particles corresponds to a
mass per particle three times larger than for the other components.
The typical number of SPH particles within $R=8$~kpc (in initial
units) and the corotation circle in the high resolution simulations is
$10^5$ and $5.5\times 10^4$ respectively.
\par The simulations are run in a completely self-consistent way,
without imposing any symmetry and taking into account all
gravitational interactions between the mass components. In particular,
the gas feels his own gravity and interacts with the stellar spiral
arms. Moreover the bar parameters are not arbitrarily chosen but
automatically and naturally adjust according to realistic dynamical
constraints. The calculations have been done on Sparc Ultra and
Silicon Graphics computers, each with over $1$~Gbyte central memory.
\begin{figure*}[p!]
~\vspace*{.5cm}\\
\centerline{\psfig{figure=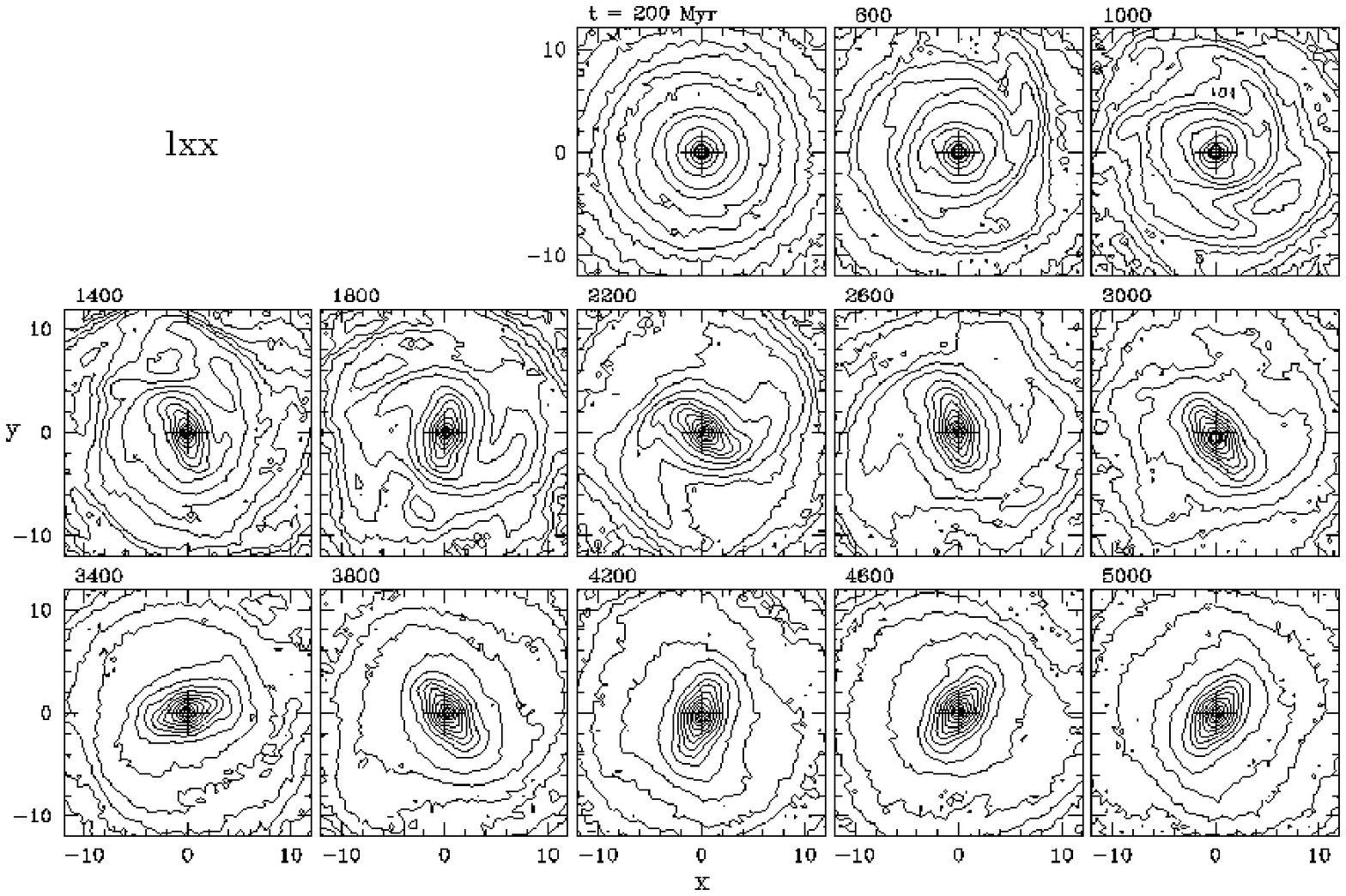,width=18cm}}
\vspace*{.5cm}\\
\centerline{\psfig{figure=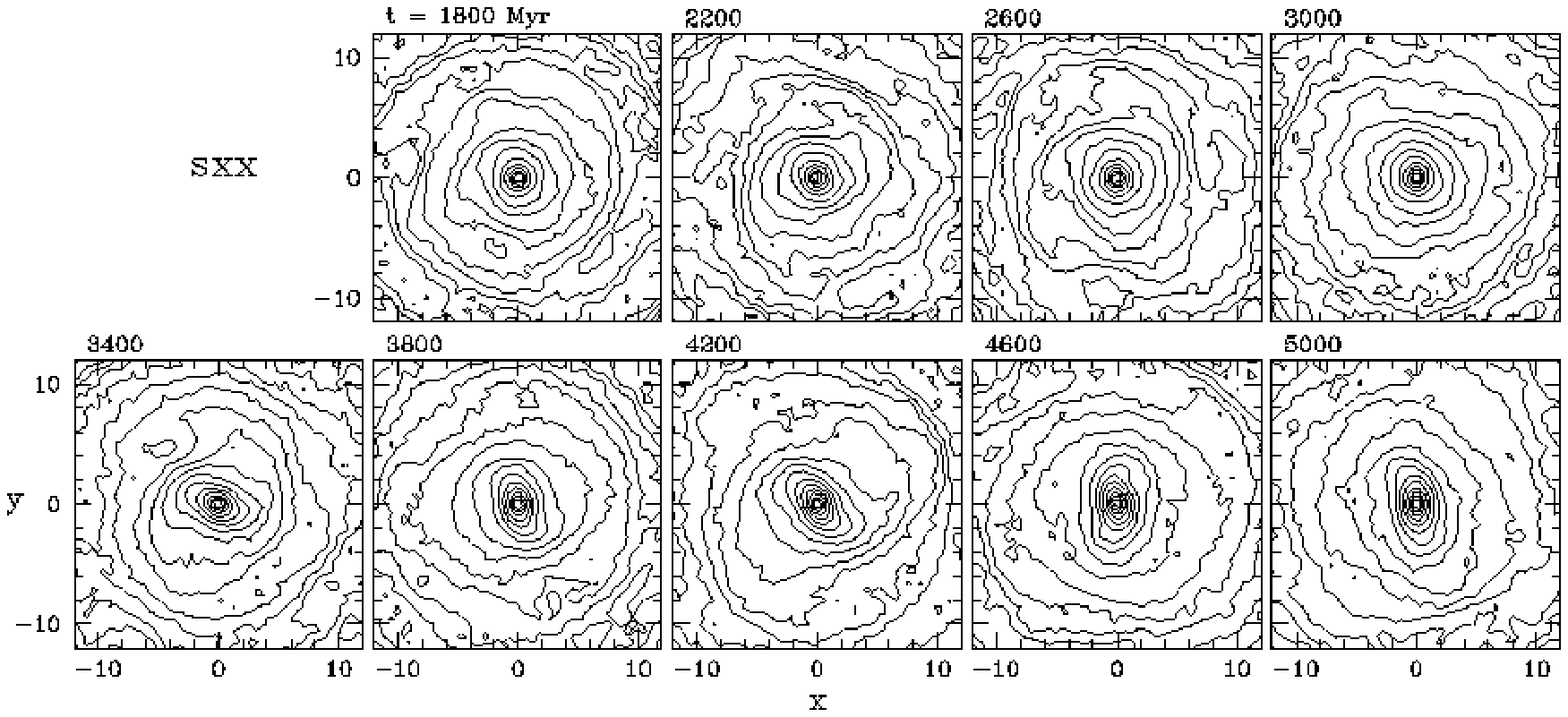,width=18cm}}
~\\
\caption{Time evolution of the face-on disc+NS surface density in
         simulations lxx and sxx with fixed gas component. The
         distances are in kpc (initial units) and the density contours
         are spaced by a constant interval of $0.5$ magnitude. The
         cross in the lxx frames indicates the position of the centre
         of mass, which coincides with the origin of the coordinates
         system.}
\label{iso}
\end{figure*}

\subsection{Stars}
\label{stars}
Figure~\ref{iso} shows the whole face-on evolution of the
simulations~lxx and~sxx with rigid gas component. Although the initial
conditions of these simulations are drawn from exactly the same
phase-space density function and the average number of particles per
grid cell is about the same, the evolution clearly depends on the
adopted resolution and number of particles: in simulation~lxx the bar
forms much more rapidly than in simulation~sxx, around $t=1.2$~Gyr
instead of $3.2$~Gyr in the latter, and is of larger extent. Moreover,
the contours of the bar become rounder close to the centre, as
observed in many external barred galaxies (e.g. M100 in the near-IR),
and the surrounding disc tends to a more flattened radial profile. An
explanation for the delay of the bar formation could be that higher
resolution simulations can catch smaller density fluctuations and
hence favour the growth rate of asymmetries. It is not clear whether
a~convergence of properties with increasing resolution has been
achieved in simulation~lxx.
\begin{figure}[t!]
\centerline{\psfig{figure=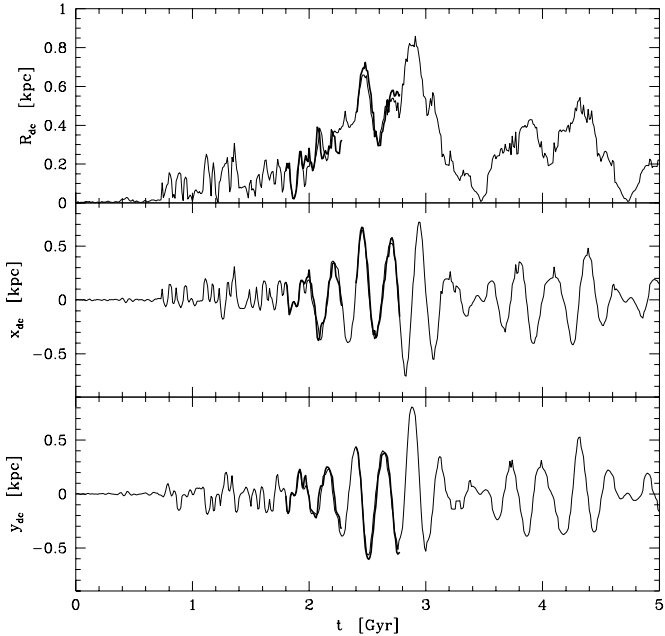,width=8.8cm}}
\caption{Radial, $x$- and $y$-displacements of the stellar density
         centre (dc) with respect to the centre of mass in simulation
         lxx (thin line) and in the live gas simulations l10 and l10'
         (thick lines).}
\label{cdd}
\end{figure}
\begin{figure}[t!]
\centerline{\psfig{figure=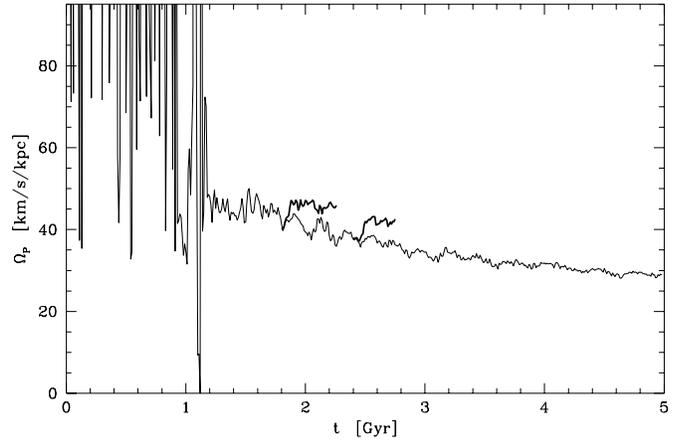,width=8.8cm}}
\caption{Pattern speed $\Omega_{\rm P}(t)$ of the bar in simulation
         lxx with fixed gas (thin line) and in simulations l10 and
         l10' with live gas (thick lines), in initial units. The
         pattern speed is derived as in paper~I by diagonalisation of
         the momentum of inertia tensor, except that the latter is
         computed relative to the offcentred density centre instead of
         the centre of mass.}
\label{omp}
\end{figure}
\par Another relevant dynamical aspect distinguishing the large
simulations from those of the s-series is the offcentring of the
stellar bar (Fig.~\ref{cdd}). At $t\approx 800$~Myr, the density
centre starts to deviate from the global centre of mass and wanders
around it. The maximum amplitude of the displacement reaches
$\sim 800$~pc at $t=2.9$~Gyr, and the revolution frequency of the
density centre amounts to $20-30$~km\,s$^{-1}$\,kpc$^{-1}$. Offcentred
bars are commonly observed in external galaxies (see Colin \&
Athanassoula 1989; Block et al. 1994; other references in Levine \&
Sparke 1998) and reported in numerical simulations of galactic discs
(e.g. Miller \& Smith 1992). At least half of all spiral galaxies have
lopsided light distribution (Schoenmakers 1999; see also Rudnick \&
Rix 1998). Figure~\ref{cdd} confirms Weinberg's~(1994) conclusion on
the persistent nature of the phenomenon. However, it might be that the
polar grids used to compute the gravitational forces amplify the
density centre offcentring, as they artificially produce an
exponential instability of the position of the centre of mass
(Pfenniger \& Friedli 1993). The same simulation repeated with a
three-dimensional Cartesian grid could result in a lower amplitude
oscillation.
\begin{figure}[t!]
\centerline{\psfig{figure=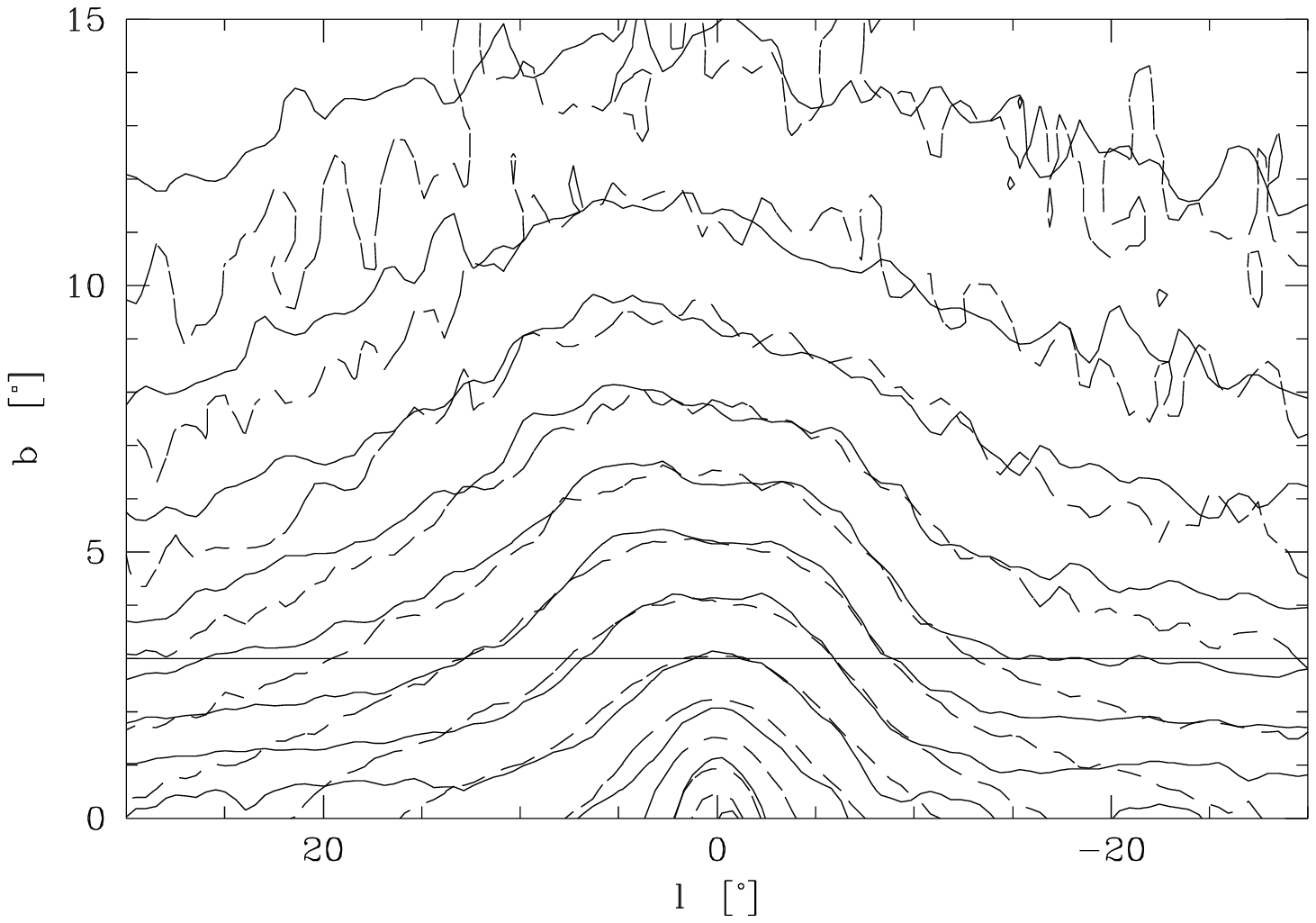,width=8.8cm}}
\caption{Result of an adjustment of model lxxt1950 to the dust
         subtracted COBE K-band image by the same technique as
         described in paper I. The solid lines show the observed
         contours spaced by half a magnitude and the dashed lines the
         corresponding model contours, assuming a constant
         mass-to-light ratio. The correction for extinction fails
         below the horizontal line.}
\label{cobe}
\end{figure}
\par The pattern speed of the bar in simulation lxx decreases roughly
exponentially from $50$~km\,s$^{-1}$\,kpc$^{-1}$ at $t=1.2$~Gyr to
$30$~km\,s$^{-1}$\,kpc$^{-1}$ at $t=5$~Gyr (see Fig.~\ref{omp}). The
face-on axis ratio $b/a$ of the bar is about $0.6$, i.e. close to the
upper limit derived from the lower resolution simulations of paper~I.
\par The phase space coordinates of the stellar particles have been
extracted from the simulations every $25$~Myr. Adjusting the location
of the observer (Sun) using the COBE/DIRBE dust subtracted K-band map
as in paper~I is complicated by the density centre offcentring: in
addition to the bar inclination angle, the relative galactocentric
distance of the observer and the mass-to-light ratio, further
parameters are needed for the direction of the Galactic centre in the
models. Here we have simply applied the standard method of paper I to
models with weakly offcentred bars, assuming that the Galactic centre
lies at the centre of mass and taking a finer Cartesian grid with
$\Delta \ell=\Delta b=0.5\degr$ to compare the data and model fluxes.
In the notations of paper I (see also Sect.~\ref{inter}) and for
$N_{\rm pix}=1200$, the best fit location parameters for model
lxxt1950 are $\tilde{R}_{\circ}=9.7$ and $\varphi_{\circ}=32\degr$,
with a mean quadratic relative residual ${\cal R}^2=0.539$\%. A
comparison of this model to the COBE data is shown in Fig.~\ref{cobe}.
\begin{figure*}[p!]
\centerline{\psfig{figure=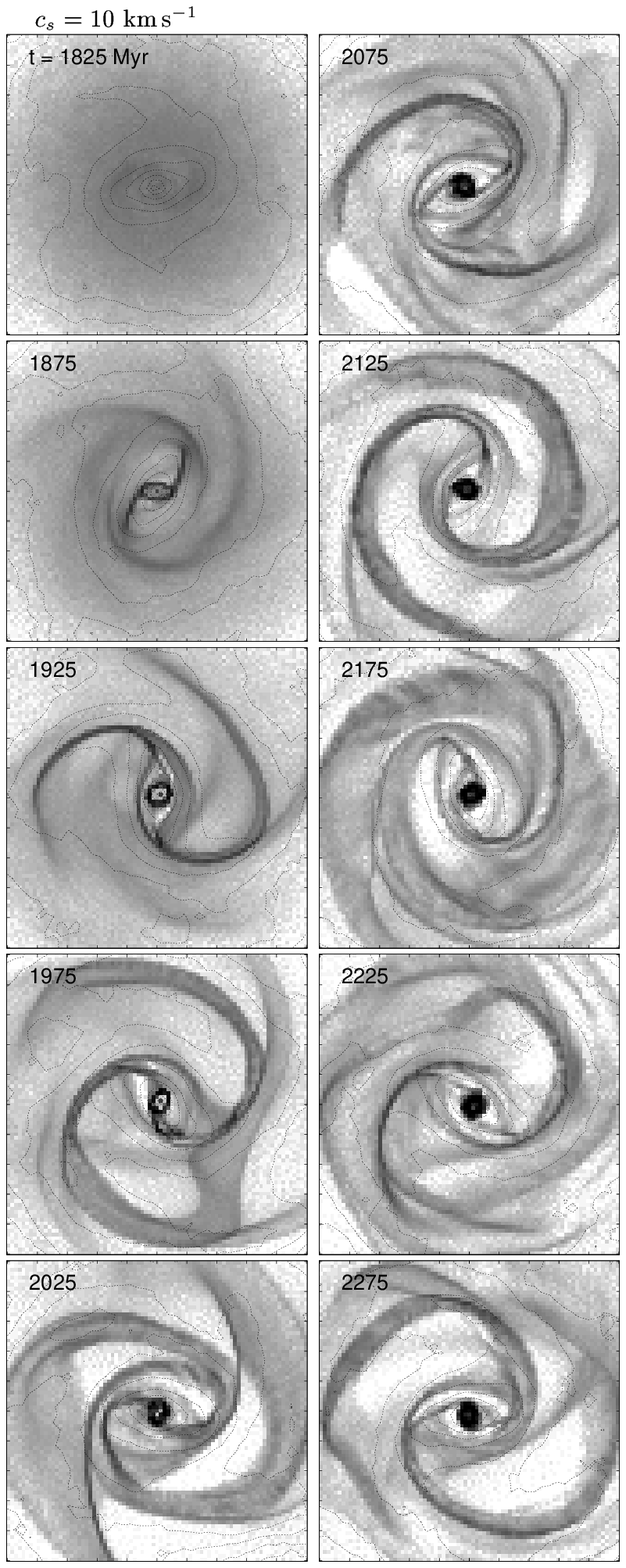,width=8.8cm}
      \hfill\psfig{figure=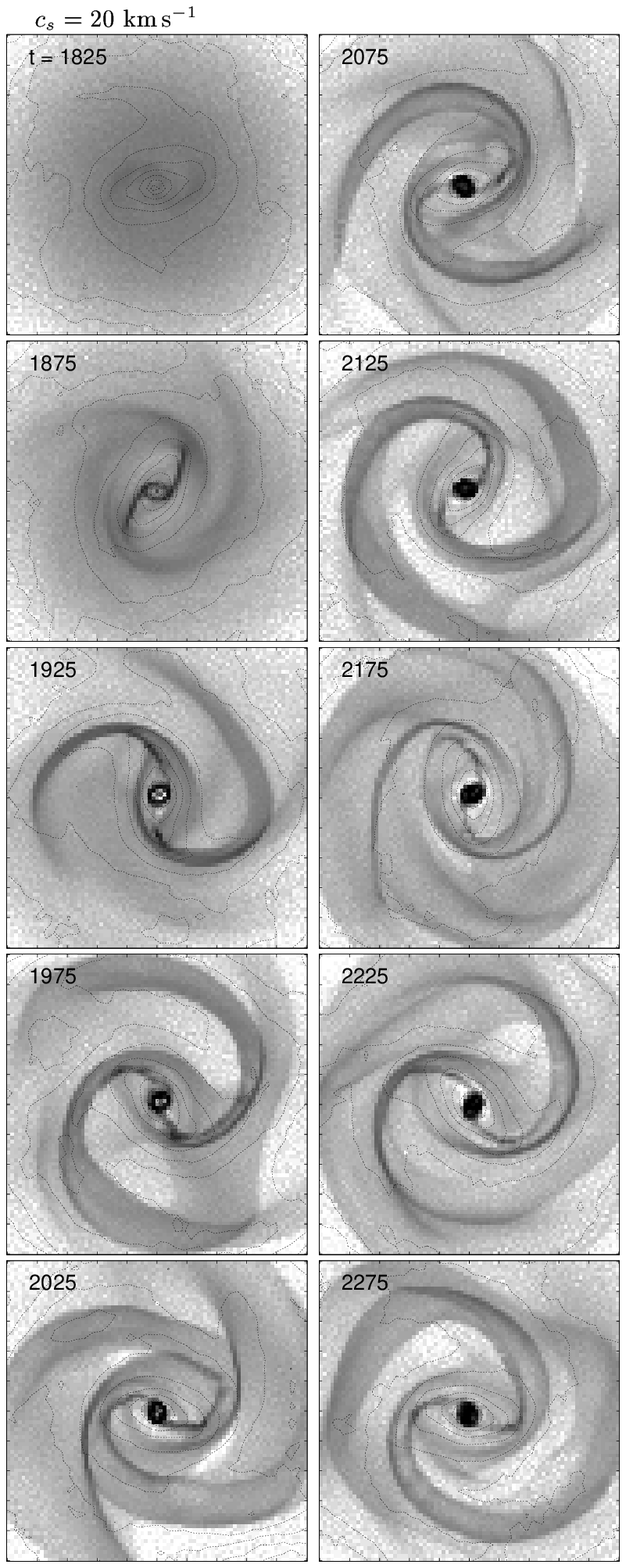,width=8.8cm}}
\centerline{\psfig{figure=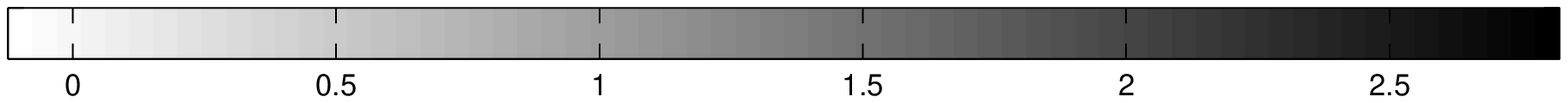,width=15.3cm}\hfill}\vspace*{-.48cm}
\centerline{\hspace*{15.1cm}
            ${\rm Log}_{10}(M_{\odot}\,{\rm pc}^{-2})$}
\caption{Face-on view of the gas flow evolution in simulations l10 and
         l20, which differ only by the value of the sound speed $c_s$.
         Each frame is $20$~kpc on a side in initial units. The dotted
         lines indicate the stellar surface mass density contours
         spaced by $0.75$ magnitude.}
\label{flow1}
\end{figure*}
\begin{figure}[t!]
\centerline{\psfig{figure=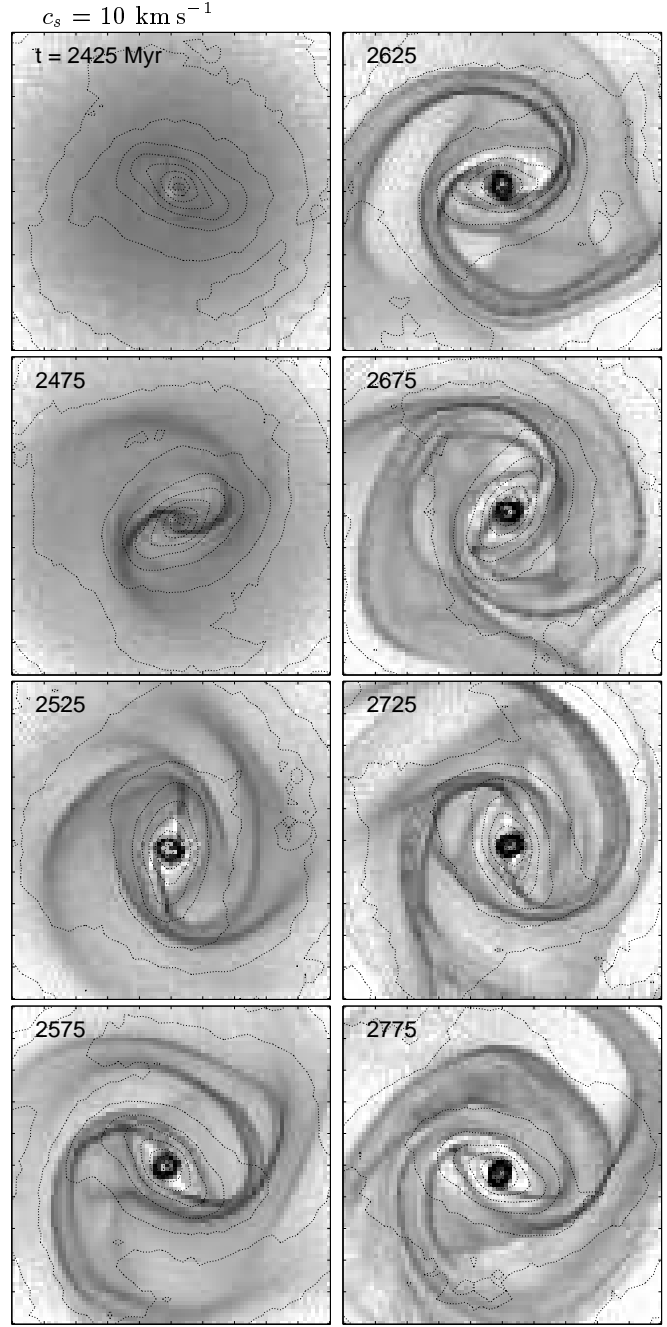,width=8.8cm}}
\caption{Face-on view of the gas flow evolution in simulation l10'.
         Each frame is $20$~kpc on a side in initial units. The gray
         scale and the dotted contours are as in Fig~\ref{flow1}.}
\label{flow2}
\end{figure}

\subsection{Gas}
\label{gas}
The time evolution of the gas flow in simulations l10, l20 and l10' is
illustrated in Figs.~\ref{flow1} and~\ref{flow2}. In each simulation
the gas flow rapidly becomes non-axisymmetric, forming transient
spiral arms, shock fronts and a nuclear ring of $x_2$ orbits
accumulated near the inner Lindblad resonance. Two kinds of spiral
structure can a priori be distinguished in the bar region (see for
instance frame $t=2075$~Myr of simulation~l10 in Fig.~\ref{flow1} or
frame $t=2775$~Myr in Fig.~\ref{flow2}):
\begin{itemize}
\item The {\it axis shocks}, or {\it off-axis shocks}, which lead more
or less the bar major axis and join the nuclear ring. These shocks,
also appearing in many other hydrodynamical simulations (e.g.
Athanassoula 1992), can be roughly understood on the basis of the
$x_1$ closed orbit family in the rotating frame of the bar (Binney et
al. 1991; Morris \& Serabyn 1996), under the approximation of a rigid
potential. Far from the centre, the gas moves along this main orbit
family because the viscous forces dissipate any libration energy
around periodic orbits. The same forces also cause the gas to switch
progressively to ever lower energy orbits and thus to approach the
centre. Below a critical value of the Hamiltonian, the $x_1$~orbits
develop loops at their apocentre where the gas dissipates some of its
streaming energy by collision. The gas then leaves these periodic
orbits to follow more radial non-periodic orbits passing round the
nuclear ring and striking the gas falling symmetrically from the other
side of the bar. The axis shocks result from the velocity difference
between the two streams, when it exceeds the sound speed. Part of the
falling gas may also collide with the central $x_2$ orbits and be
directly absorbed by the nuclear ring. Axis shocks have been detected
in the velocity field of several external barred galaxies (see
Sect.~\ref{ban}) and seem to be associated with the prominent
dustlanes leading the bar in these galaxies. The dust grains are
strongly concentrated behind the shock fronts and thus produce the
typical extinction signatures. Some prototypical galaxies with offset
dustlanes are NGC~1300, NGC~1433, NGC~1512, NGC~1530, NGC~1365 and
NGC~6951, which are all SBb or SBbc type galaxies (see e.g. Sandage \&
Bedke 1988), as the Milky~Way. The self-gravity of the gas certainly
plays an important role in holding together the shocked gas. 
\item The {\it lateral arms}, which roughly link the bar ends avoiding
the nuclear ring by a large bow, and generally correspond to the inner
prolongation of spiral arms in the disc. The gas moves almost parallel
to these arms and finally meets the axis shocks. According to Mulder
\& Liem~(1986), the \mbox{3-kpc} arm is almost certainly of this kind.
When a lateral arm forms, the part close to the axis shock where the
gas runs into is located well inside the bar. Then the arm moves
outwards along the axis shock and progressively dissolves as it
approaches corotation (this will be illustrated more quantitatively in
Sect.~\ref{con}). Such dissolution may be partly linked with the
decreasing gravitational resolution with radius in the simulations.
The gaseous lateral arms in our simulations resemble the innermost,
sometimes ring-like, stellar arms in external barred galaxies, e.g.
NGC~1433, NGC~4593, NGC~6951, NGC~3485, NGC~5921, NGC~7421, which are
also SBb or SBbc type galaxies (Sandage \& Bedke 1988).
\end{itemize}
This classification should only be considered as a first order guide.
The main distinction between the two structures is that the axis
shocks intersect or brush the nuclear ring, while~\pagebreak the
lateral arms pass away from it.
\begin{figure*}[t!]
\centerline{\psfig{figure=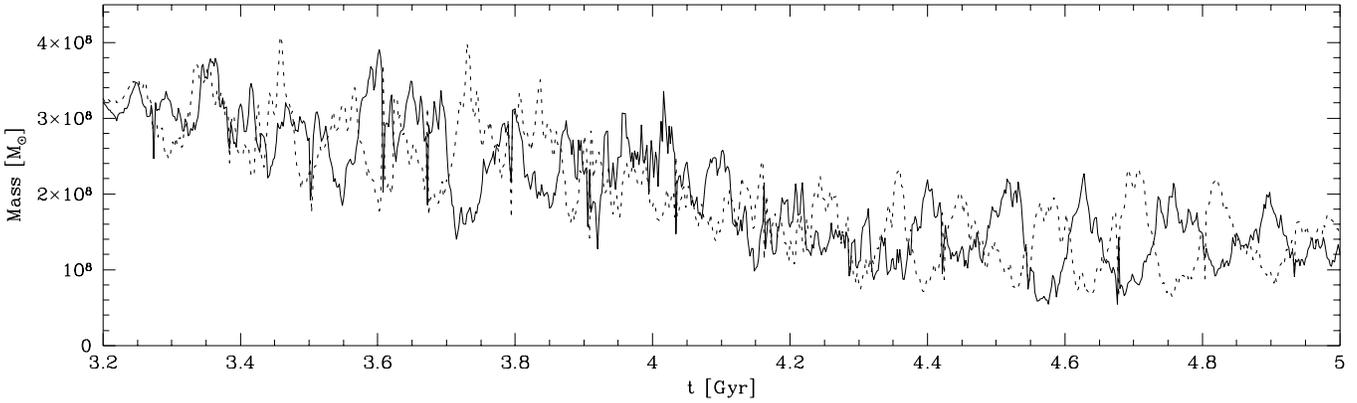,width=18cm}}
\caption{Asymmetric alternated cycle of the gas flow on the bar flanks
         in the small simulation~s10. The full and dotted lines
         indicate the gas~mass inside each of the two regions
         corotating with the bar, located within $3$~kpc from the
         centre and at least $500$~pc away from the bar major axis.
         The cycle is especially marked in the last $700$~Myr of the
         simulation (in which the bar never gets significantly
         offcentred).}
\label{cycle}
\end{figure*}
\par Increasing the sound speed, i.e. the pressure forces relative to
the gravitational forces, yields as expected a smoother gas
distribution and softens the sharp corners in the spiral arms (see
Fig.~\ref{flow1}). The nuclear ring slightly shrinks and the axis
shocks occur closer to the major axis of the bar, in agreement with
the SPH results of Englmaier \& Gerhard~(1997). Note that some
sub-structures also seem to develop within the nuclear ring for
$c_s=20$~km\,s$^{-1}$.
\par The gas flow never reaches a stationary state and is most of the
time asymmetric, as is often observed in external disc galaxies. In
particular, the lateral arms are rarely symmetric both in position and
surface density, and the axis shocks considerably vary in shape.
Sometimes, only one lateral arm or axis shock can be distinguished,
and sometimes these structures are doubled, appearing twice on the
same side of the bar (see e.g. l10t2175 in Fig.~\ref{flow1} for double
lateral arms). It may even happen, as is the case in frame
$t=1975$~Myr of simulation~l10 (Fig.~\ref{flow1}), that the axis
shocks receive a sufficient impulse from the lateral arms to transform
themselves into such arms. The nuclear ring, owing to its
gravitational coupling to the stellar components, intimately follows
the bar density centre in its oscillations around the centre of mass.
The inclusion of gas prevents the pattern speed of the bar to slow
down (Fig.~\ref{omp}; see also Friedli \& Benz 1993; Berentzen et al.
1998): after a substantial initial readjustment, $\Omega_{\rm P}$
remains very nearly constant over the short duration of the live gas
simulations.
\par In the s-series, the moderate bar considerably weakens once the
gas is switched-on. However, in simulation~s10 it survives for a
sufficiently long time to reveal an interesting asymmetric process in
the gas flow (Fig.~\ref{cycle}): the gas seems to rarefy alternatively
on each side of the bar intermediate axis according to a cycle of
roughly $125$~Myr. As a large amount of gas flows on one side, the
other appears almost devoided of gas, and when the gas of the former
side reaches the axis shock, the situation reverses and the previously
depleted side is filled. During this cycle, the gas density on the bar
flanks can vary from single to triple.
\par The non-stationarity of the gas flow in the l-simulations is not
a simple unachieved readjustment due to non-equilibrium in the initial
conditions and the rather short integration time, since the smaller
simulations with live gas, integrated much longer in time, themselves
always remain time-dependent. Outside the bar region, the stellar
spiral arms have a differential pattern speed, in general lower than
the bar (see e.g. Sellwood \& Sparke 1988), and drive the gaseous arms
which repeatedly wound and dissolve. Many SPH simulations carried in
rigid barred potentials seem to finally settle in a stationary spiral
structure rotating at the same angular speed as the bar. Such results
are an artificial consequence of the imposed fixed background
potential.

\section{Interpretation of the $\ell-V$ features}
\label{inter}
Model $\ell-V$ diagrams depend on the location of the observer in the
galactic plane, which is described by his galactocentric distance
$\tilde{R}_{\circ}$ in initial units and the angle $\varphi_{\circ}$
between the line joining himself to the centre and the major axis of
the bar, and on his velocity $\vec{v}_{\circ}$. We assume here that
the observer lies in the plane $z=0$ and that $\vec{v}_{\circ}$ is
purely azimuthal.
\par The phase space coordinates of the SPH particles have been stored
every $2$~Myr in order to realise a posteriori $\ell-V$ diagram movies
for any arbitrary values of these parameters with a high temporal
resolution. To compare these movies with the observations and decide
for an optimum model, some reference points have been overlayed on the
movies to localise the features seen in the CO and HI data
(Table~\ref{rep}). In each movie, the view point is at rest relative
to the rotating frame of the bar, i.e. $\tilde{R}_{\circ}$ and
$\varphi_{\circ}$ remain constant. The contribution of each SPH
particle to the $\ell-V$ frames is weighted by its inverse squared
distance relative to the observer to mimic the flux decline of point
sources (all model $\ell-V$ diagrams in this paper are computed this
way, except the one in Fig.~\ref{x1}). A direct consequence of the
asymmetries developing in our simulations is that model $\ell-V$
diagrams computed for diametrically opposite view points always
considerably differ.
\begin{table}[t!]
\centering
\caption{Reference points overlayed in the $\ell-\!V$ diagram movies
         to locate the main observed features indicated in
         Fig.~\ref{lv} and the Carina~arm.}
\vspace*{.2cm}
\begin{tabular}{lrr}
Reference & $\ell$ [$\degr$] & \hspace{-.2cm}$V$ [km\,s$^{-1}$]\hspace{-.2cm}
                                                                    \\ \hline
Positive bright velocity peak                     &   $3$ &  $270$~ \\
Negative bright velocity peak                     &  $-2$ & $-220$~ \\
Intersection of the 3-kpc arm with the $\ell$-axis & $12$ &    $0$~ \\
Knee of the 3-kpc arm                             & $-17$ & $-125$~ \\
Knee of the 135-km\,s$^{-1}$ arm (Clump 1)        &  $-5$ &  $100$~ \\
Maximum emission of the connecting arm     & $7.5$\hspace{-.25cm} &  $150$~ \\
Tangent points of the molecular ring branches     &  $25$ &  $120$~ \\
                                                  &  $30$ &  $115$~ \\
Tangent point of the Carina arm                   & $-78$ &    $0$~ \\ \hline
\label{rep}
\end{tabular}
\end{table}
\par Two models from these movies, l10t2066 and l10't2540, illustrated
and confronted to the observations in Fig.~\ref{models}, have
especially retained our attention. In both cases, the inclination
angle of the bar is $25\degr$ and no velocity rescaling has been
applied. The circular velocity of the observer is set to the mean
azimuthal velocity of his surrounding SPH particles, i.e. to
$203$~km\,s$^{-1}$ for l10t2066 and $197$~km\,s$^{-1}$ for l10't2540.
The model l10't2540 is selected as one of the models which best
reproduce the overall $\ell-V$ observations within the solar circle.
Even if the agreement is rather qualitative, this model offers a solid
basis to interpret the data. Figure~\ref{region} highlights the spiral
arms which trace the $\ell-V$ features in the model and
Fig.~\ref{vfield} shows the velocity field of the gas.
Figure~\ref{locate} also provides a qualitative key-map to spatially
locate observed $\ell-V$ sources in the Galactic plane.

\subsection{Connecting arm}
\begin{figure*}[p!]
\centerline{\psfig{figure=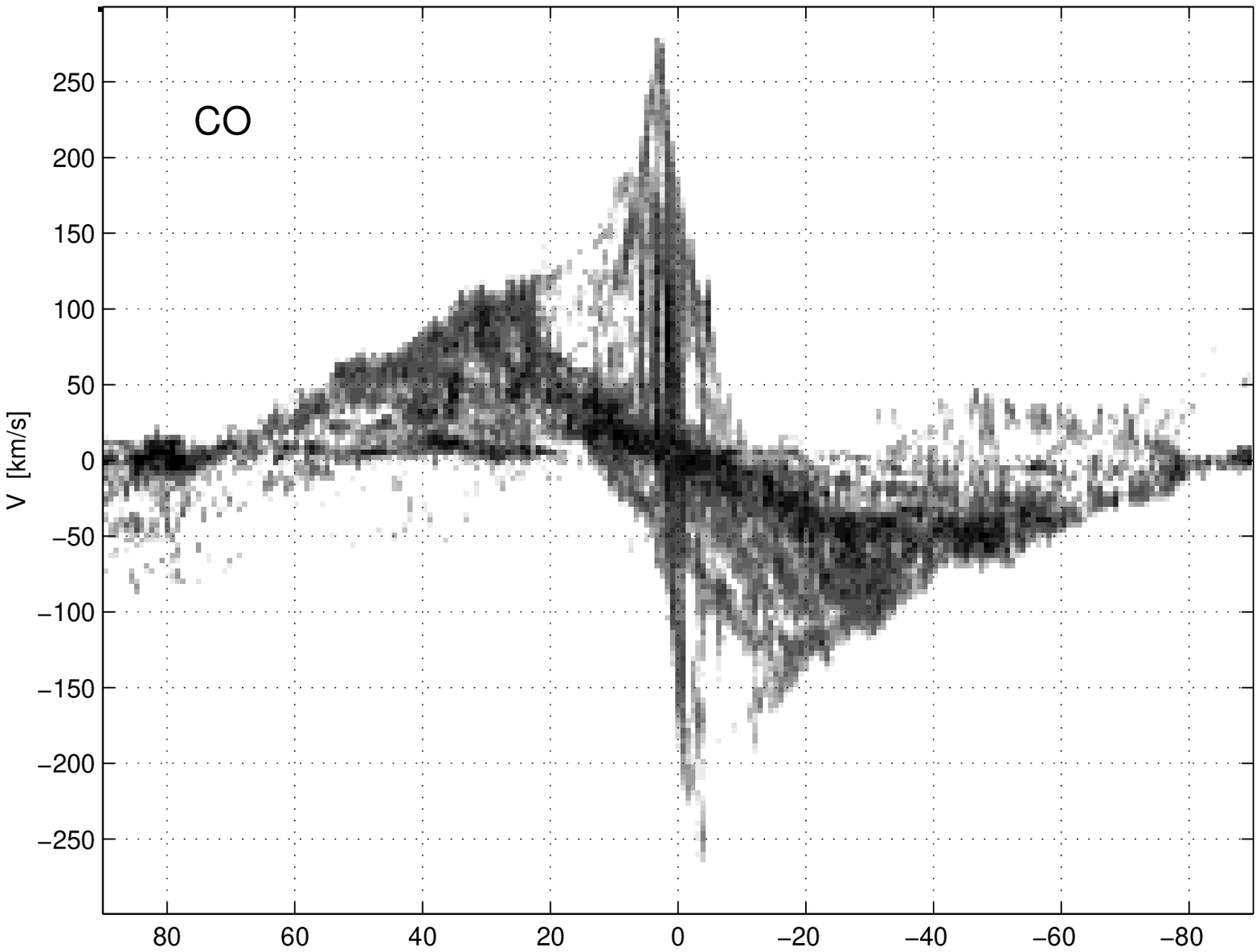,width=8.75cm}
            \psfig{figure=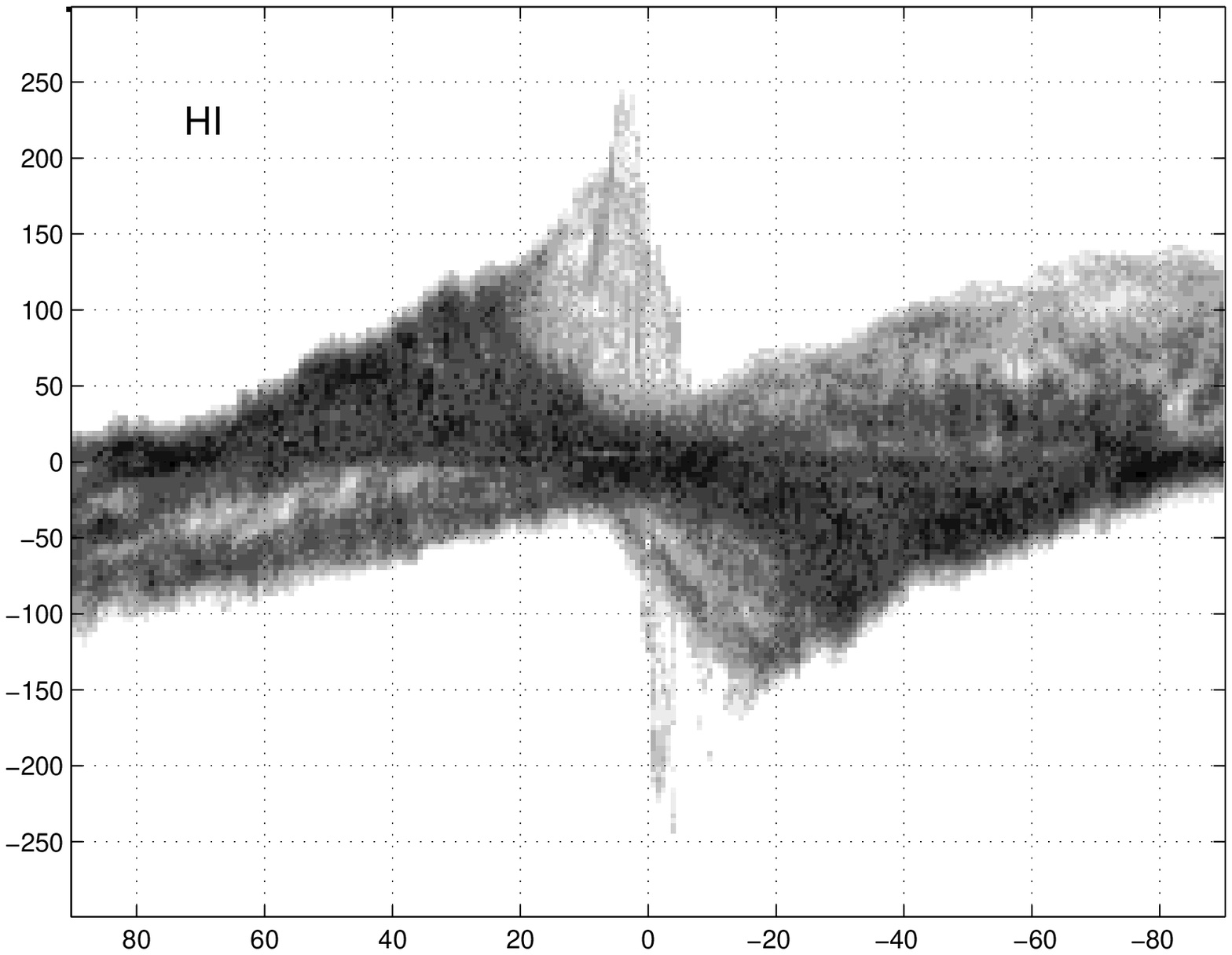,width=8.75cm}}
\centerline{\psfig{figure=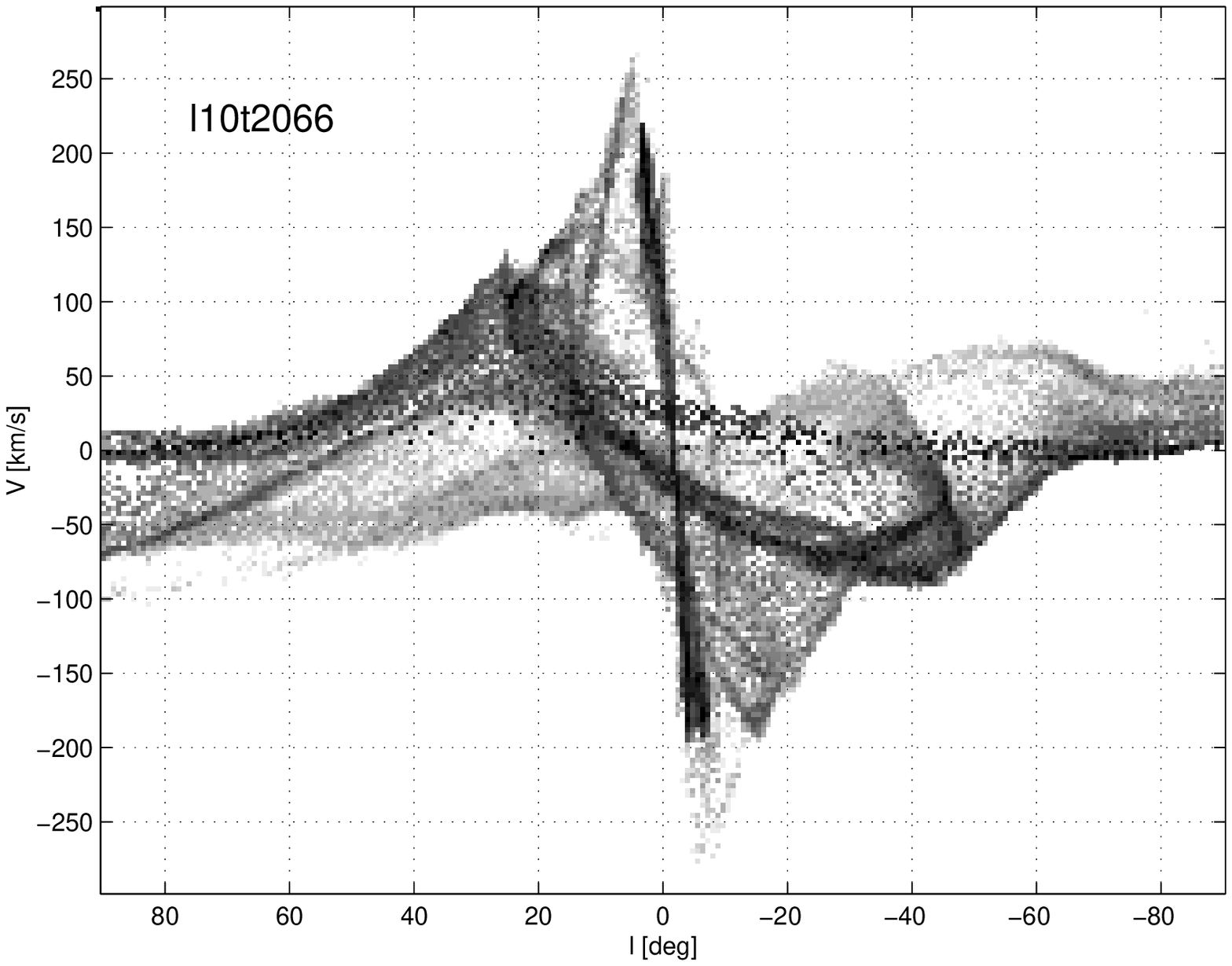,width=8.75cm}
            \psfig{figure=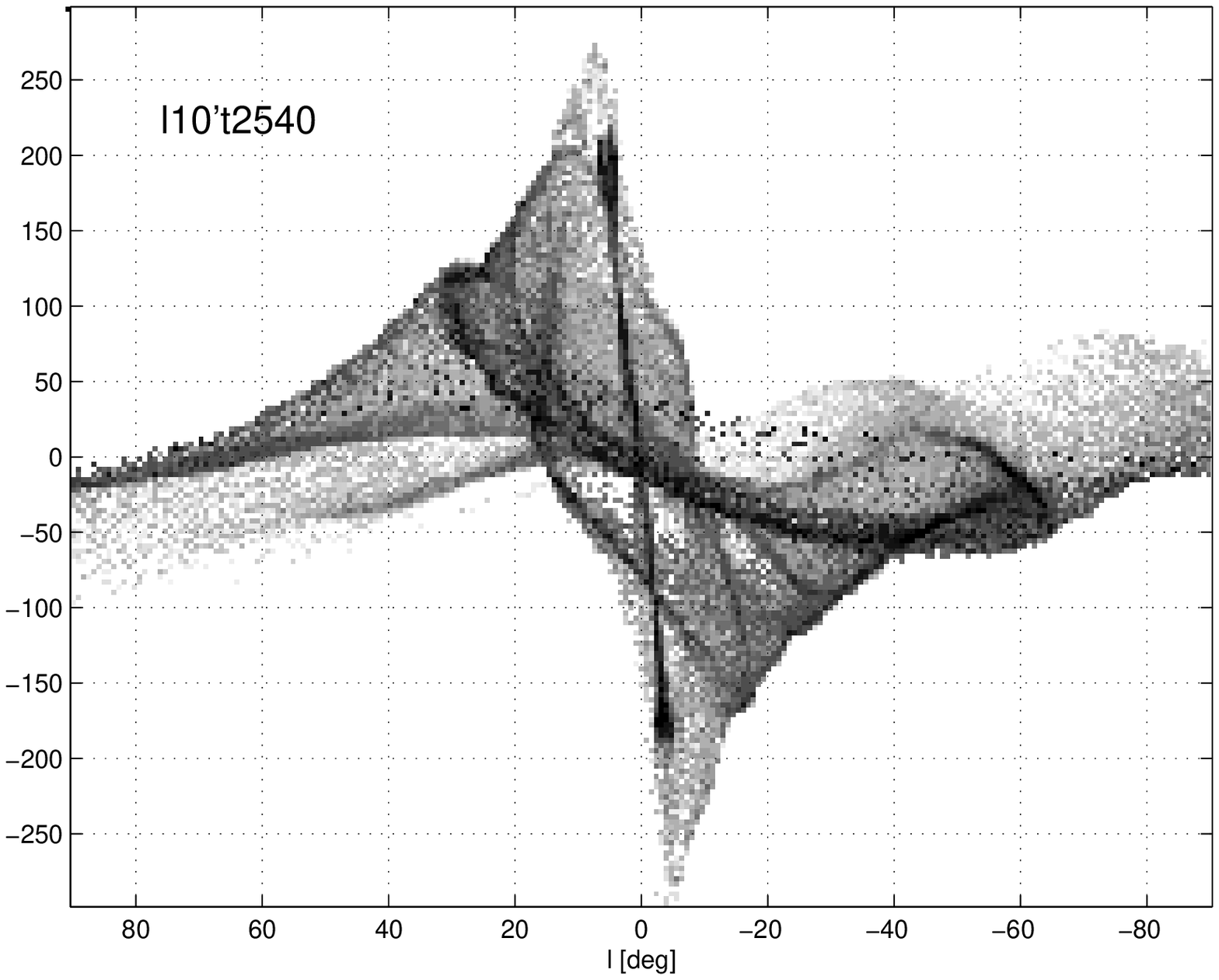,width=8.75cm}}
\centerline{\hspace*{.7cm}\psfig{figure=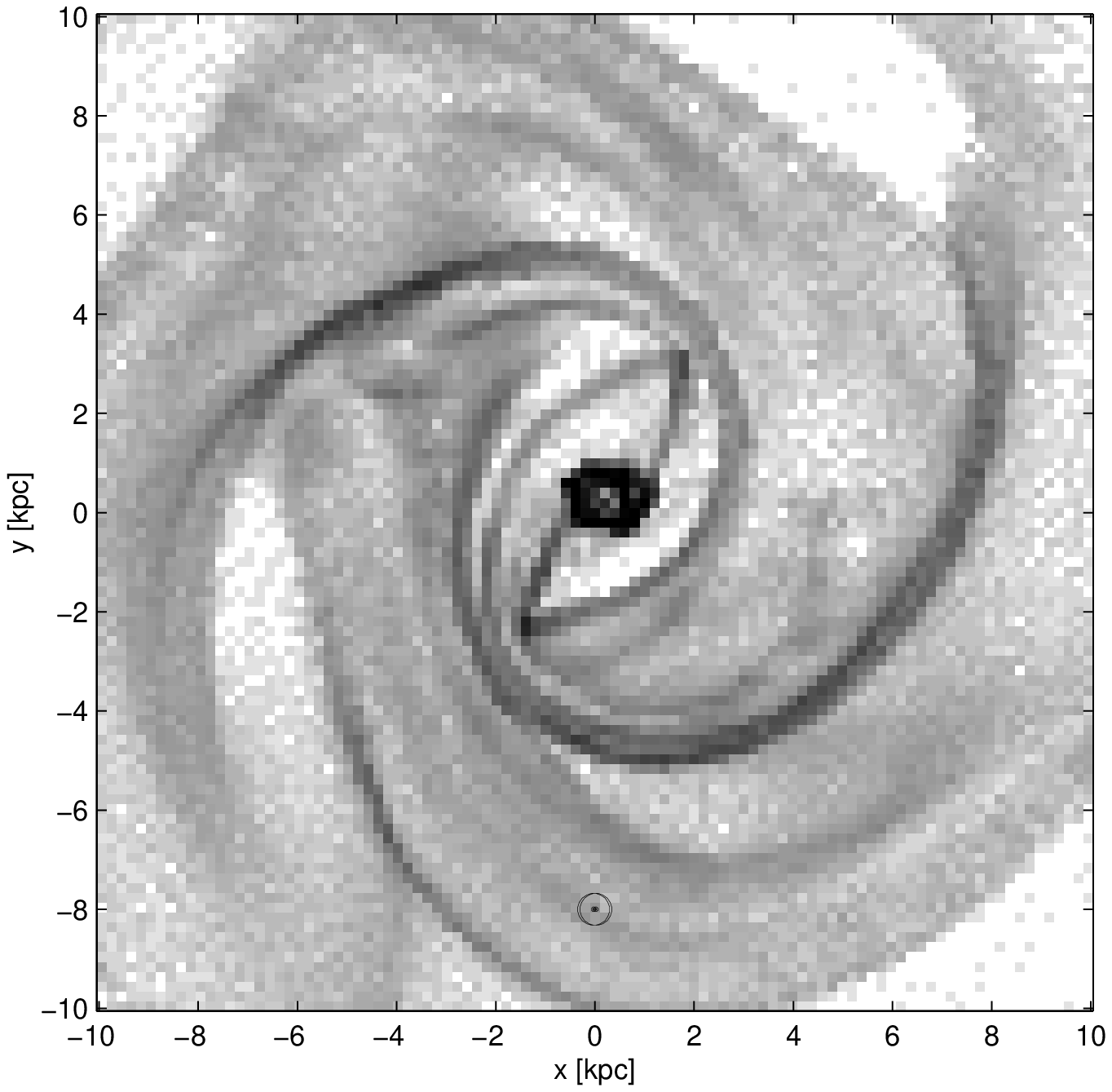,width=8.75cm}
                          \psfig{figure=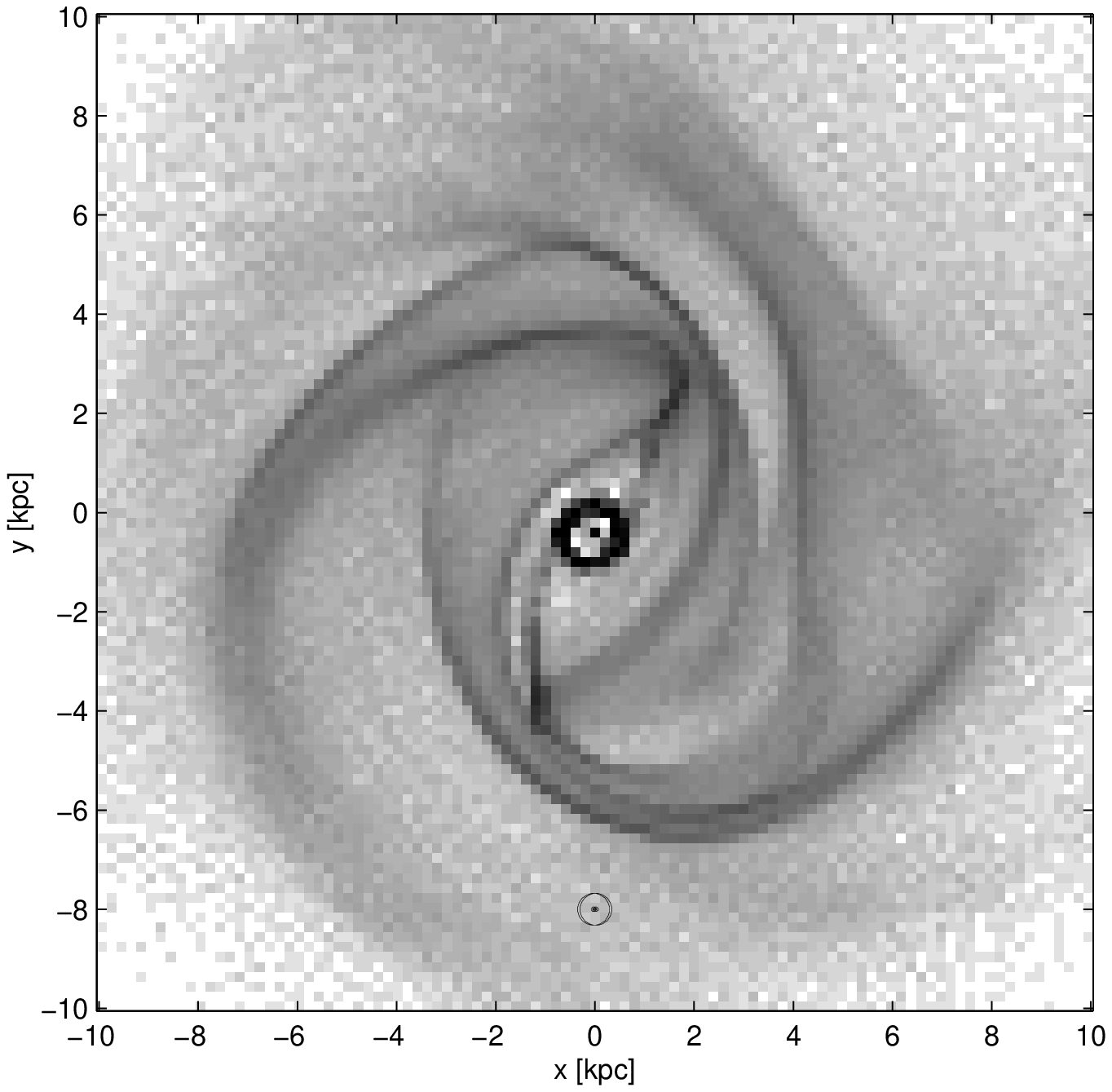,width=8.75cm}}
\caption{Confrontation of a selection of two models with observed gas
         kinematics. Top: $^{12}$CO and HI $\ell-V$ diagrams
         integrated over $|b|\leq 2\degr$ and $|b|<1.25\degr$
         respectively; the data are from Dame et al.~(1999) for the
         CO, and Hartmann \& Burton~(1997), Burton \& Liszt~(1978) and
         Kerr et al.~(1986) for the HI. Middle: synthetic $\ell-V$
         diagrams of models l10t2066 and l10't2540 for a bar
         inclination angle $\varphi_{\circ}=25\degr$, including all
         particles within $|b|<2\degr$. Bottom: face-on projections of
         the gas spatial distribution in these models, rescaled such
         as to put the observer at $(x,y)=(0,-8)$~kpc ($\odot$
         symbol). In these units, corotation lies at
         $R_{\rm L}=4.5$~kpc (l10t2066) and $4.4$~kpc (l10't2540). The
         model on the left reproduces almost perfectly the connecting
         arm, while the model on the right provides a fair global
         qualitative agreement to the data.}
\label{models}
\end{figure*}

%
The connecting arm is clearly identified with the axis shock in the
near part of the bar. More precisely, this arm is build up by the gas
clouds which have recently crossed the shock front at various
galactocentric distances and which now collectively plunge towards the
nuclear ring/disc with velocities roughly parallel to the shock front.
In other words, the connecting arm traces the near-side branch of the
Milky~Way's dustlanes. Other models, like l10t2066 in
Fig.~\ref{models}, exhibit axis shocks with $\ell-V$ traces resembling
much more the real connecting arm than in model l10't2540, thus
reinforcing our interpretation.
\par The presence of the connecting arm feature in the observed
$\ell-V$ diagrams can be considered as a further evidence of the
Galactic bar. Furthermore, its rather large domain in longitude is
relevant of an extended bar, especially if the latter is seen close
from its major axis (see Sect.~\ref{con}). The connecting arm feature
is a real concentration of gas in space and not an artifact due to
velocity crowding along the line of sight, as suggested by Mulder \&
Liem~(1986)\footnote{These authors pretend that the connecting arm
intervene between longitudes $10\degr$ and $20\degr$, which obviously
conflicts with the sources (van der Kruit 1970 and Cohen 1975) they
mention.}. In our simulations, the gas does not always trace the full
length of the axis shocks and hence it is not surprising that the
emission from the connecting arm appears truncated at
$\ell\ga 10\degr$. The $\ell-V$ movies sometimes show gas lumps
deposited by lateral arms in the near-side axis shock and moving
precisely along the connecting arm feature. The time for the gas to
travel from the endpoint of a lateral arm to the nuclear ring along
the shock is of order $15-20$~Myr and it takes about another $10$~Myr
for the gas not absorbed by the ring to encounter the opposite axis
shock. The fraction of gas deposited in the nuclear ring is
time-dependent, but has been estimated to $20$\% in a steady gas flow
model of NGC~1530 (Regan et al. 1997). The gas mass in each branch of
the axis shocks is $\sim 4\times 10^7$~$M_{\odot}$ or less.
\par
\begin{figure}[b!]
\centerline{\psfig{figure=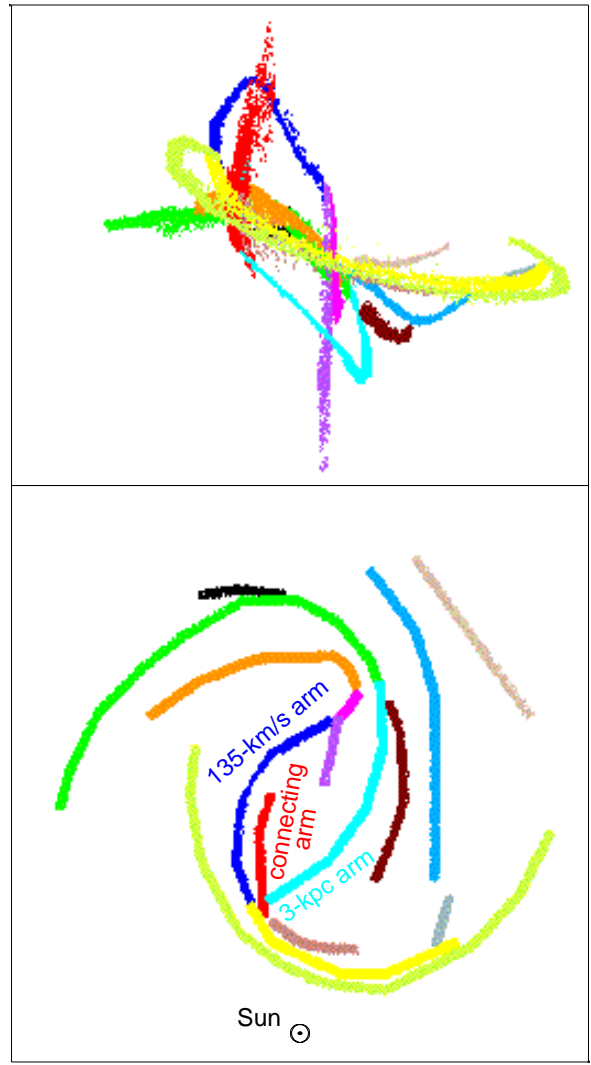,width=7.8cm}}
\caption{Link between the spiral arms in the $x-y$ plane and the
         $\ell-V$ features in model l10't2540. The spiral arms and
         their $\ell-V$ traces are depicted by the true phase space
         coordinates of the gas particles, using all particles within
         a narrow band of $300$~pc centred on the maximum surface
         density curve of each spiral arm. In the upper plot, closer
         structures relative to the observer have been overlayed on
         the more distant ones. The nuclear ring is not represented.}
\label{region}
\end{figure}

 According to Fig.~\ref{region}, the axis shock in the far-side of
the bar is predicted as an almost vertical feature in the $\ell-V$
diagram, i.e. with $\ell\approx$ constant. Such a feature is indeed
visible in the $^{12}$CO observations (Fig.~\ref{lv}) near
$\ell=-4\degr$, with only a marginal decrease of absolute longitude
towards negative velocities. The longitude confinement comes from the
fact that the shock line is nearly parallel to the line of sight and
the velocity extension from the fact that the velocity of the gas
along the shock rapidly increases when approaching the nuclear
ring/disc.

\subsection{3-kpc and 135-km\,s$^{-1}$ arms}
\label{lat}
\begin{figure}[b!]
\centerline{\psfig{figure=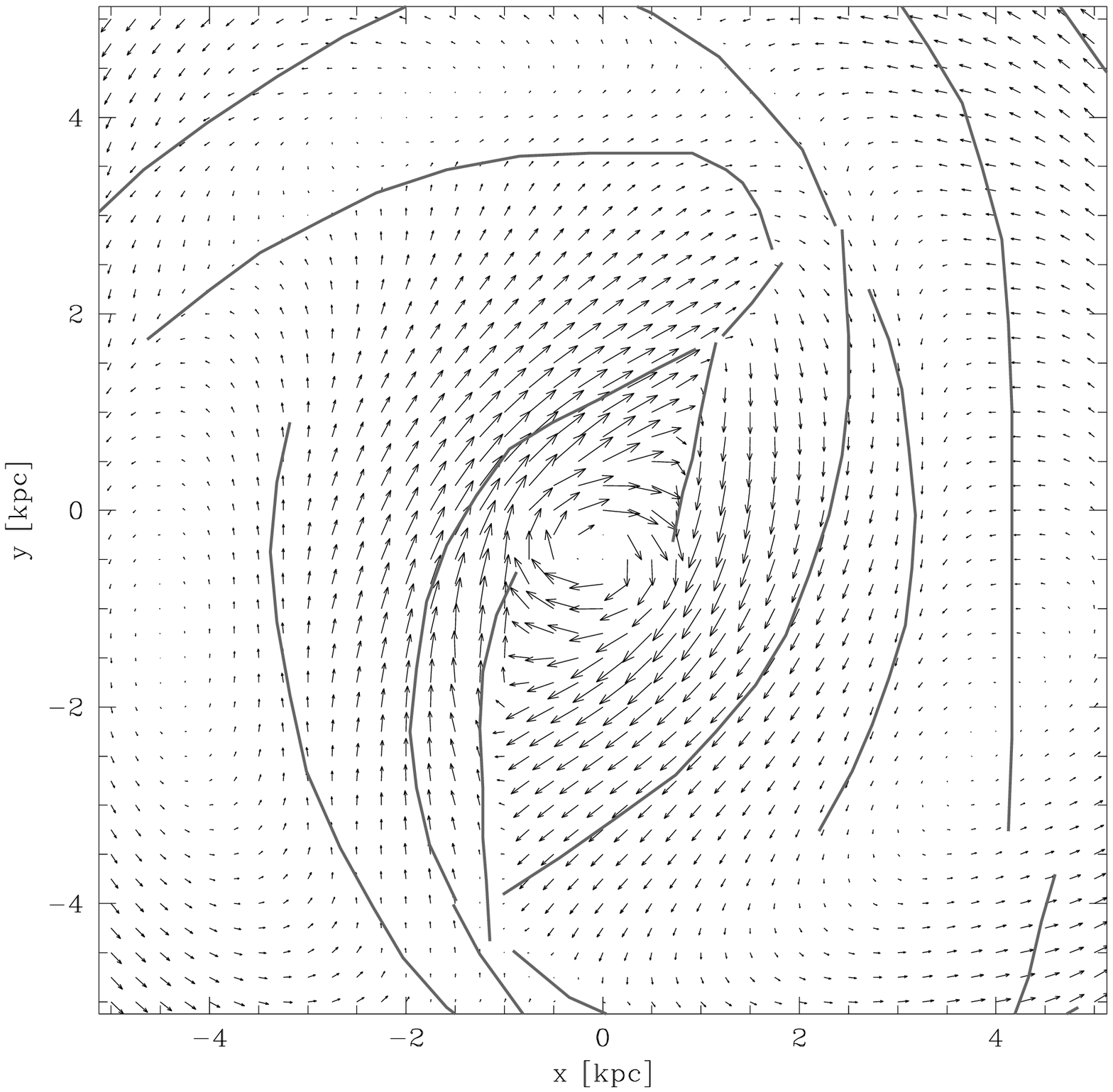,width=8.8cm}}
\caption{Velocity field of model l10't2540 in the rotating frame of
         the bar. The subtracted solid rotation does not take into
         account the bar offcentring. The grey curves indicate the
         location of the spiral arms.}
\label{vfield}
\end{figure}

The 3-kpc and the 135-km\,s$^{-1}$ arms are lateral arms. The observed
velocity asymmetry between these two arms happens because the latter
passes closer to the centre than the former. The gas associated with
the 135-km\,s$^{-1}$ arm, moving almost parallel to the arm
(Fig.~\ref{vfield}), indeed falls deeper in the potential well of the
nucleus and therefore reaches higher forbidden velocities before
striking the dustlane shock.
\par The observed velocity asymmetry of bright emission near the
positive and negative terminal velocity peaks could have a related
origin: the gas from the 3-kpc arm, after crossing the dustlane shock
further out than its counterarm, starts to fall towards the nuclear
ring/disc with a higher potential energy and therefore will approach
the latter with higher velocities, producing an enhanced velocity
peak. Model l10't2540 however does not exhibit such a velocity
asymmetry. This asymmetry could also arise from a relative radial
motion between the nuclear ring/disc and the LSR, produced either by
the oscillations of the density centre (see Sect.~\ref{stars}) or by a
radial velocity component of the LSR with respect to the Galactic
centre, possibly induced by the bar itself (Raboud et al. 1998), or
both.
\par The inner branch of the molecular ring, with tangent point at
$\ell\approx 25\degr$, is the outer extension of the 135-km\,s$^{-1}$
arm.

\subsection{Bania's clumps}
\label{ban}
Observations of the gas velocity field in external barred galaxies
have revealed velocity changes up to $200$ km\,s$^{-1}$ across the bar
leading dustlanes, demonstrating that these dustlanes are associated
with very strong shocks. Such velocity fields have been measured for
example in NGC~6221 (Pence \& Blackman 1984), NGC~1365 (Joersaeter
\& van Moorsel 1995; Lindblad et al. 1996), NGC~3095 (Weiner et al.
1993), NGC~1530 (Regan et al. 1997; Reynaud \& Downes 1998) and
NGC~7479 (Laine et al. 1999). Figure~\ref{shock} shows how the gas
velocity field is affected by the near-side branch of the axis shocks
in model~l10t2066. The shock occurs near $x=-0.55$~kpc and is followed
downstream by a huge gas concentration. Both velocity components,
parallel and perpendicular to the shock front, undergo an abrupt
velocity gradient across the shock layer. The velocity change is
however larger for the parallel component, reflecting the shearing
nature of the axis shocks, and is comparable to the values observed in
external galaxies. In NGC~1530, Reynaud \& Downes~(1998) found that
the velocity change increases towards the nuclear ring, but this
property is hard to infer from our models because the shock fronts are
not well resolved in the low density part.
\par
\begin{figure}[t!]
\centerline{\psfig{figure=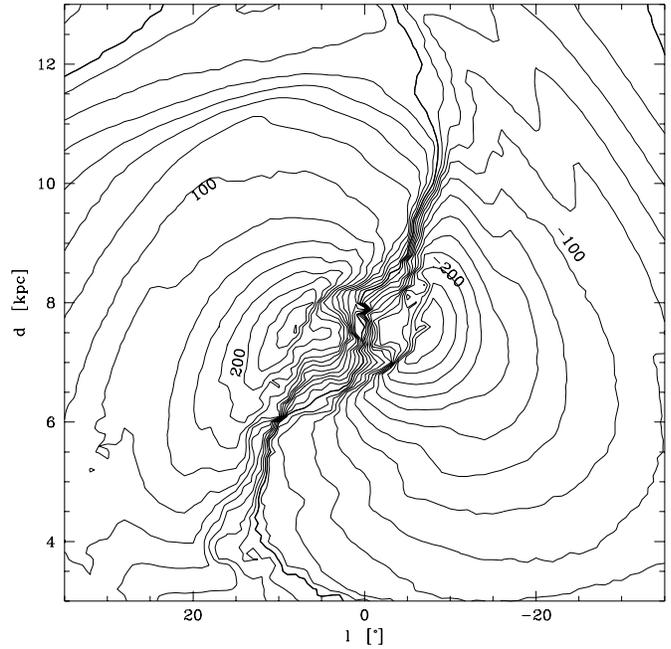,width=8.8cm}}
\caption{Radial velocity contours in model l10't2540 as a function of
         longitude and distance relative to the observer, whose
         location and circular velocity are as in Fig.~\ref{models}.
         The spacing of the contours is $20$~km\,s$^{-1}$ and the
         thick contour is the zero velocity curve (labels are given in
         km\,s$^{-1}$). Only the particles within $|b|<2\degr$ have
         been considered. Note the very sharp velocity gradient when
         crossing the axis~shocks.}
\label{locate}
\end{figure}

\begin{figure}[t!]
\centerline{\psfig{figure=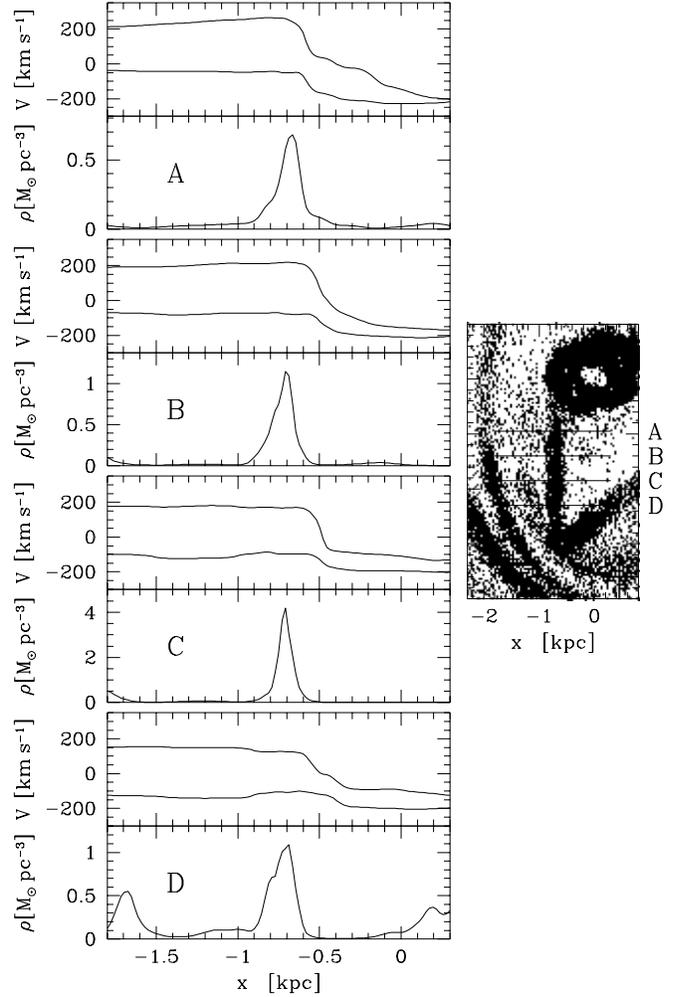,width=8.8cm}}
\caption{Spatial density and velocity profiles along four slits
         perpendicular to the ``connecting arm'' in model l10t2066.
         The upper and lower velocity curves respectively stand for
         the velocity components parallel ($V_y$) and perpendicular
         ($V_x$) to the shock front. The profiles are derived by
         standard SPH summation, e.g. from Eq.~(\ref{rho}) for $\rho$,
         and for $z=0$. The vertical distribution of the dense
         post-shock gas everywhere peaks within $10$~pc from that
         plane.}
\label{shock}
\end{figure}

 Bania's~(1977) clump 2~and an other vertical feature near
$\ell \approx 5.5\degr$ (see Fig.~\ref{lv}) could represent gas lumps
which are just about to cross the near-side dustlane shock somewhere
between the 3-kpc arm and the nuclear ring/disc: the upstream part
still moves with the small quasi-apocentric velocities of the
pre-shock orbits, while the downstream part has been accelerated to
the high inwards post-shock velocities, giving rise to a steep radial
velocity gradient (Fig.~\ref{locate}) and a velocity stretch of over
$100$~km\,s$^{-1}$ in the observations. Contrary to the $\ell-V$ trace
of the far-side dustlane shock, these features might be really
concentrated in space and not result from an accumulation of gas along
the line of sight. For the $\ell=5.5\degr$ lump, this interpretation
is supported by the fact that all emission from the lump originates at
nearly constant latitude, as expected from a spatially confined
source, and that the part of the connecting arm at the same longitude
as the lump appears at almost the same latitude as the lump itself
(see Fig.~\ref{lvb}). Furthermore, this lump also has a small mass
relative to the connecting arm and will therefore essentially adapt
its momentum to that of this arm. For clump~2, with a mean latitude
differing by more than $0.5\degr$ from that of the connecting arm (at
similar $\ell$) and a total mass of nearly $10^7$~$M_{\odot}$ (Stark
\& Bania 1986), our interpretation is more speculative. However, if
this clump is indeed close to the apocentre of its orbit, it will
enter very slowly the shock line, where gas is moving at very high
speed (over $200$~km\,s$^{-1}$), and thus receive a significant
impulse when integrated over time. Moreover, the clump complex may
move on a kind of looped quasi-$x_1$ orbit and therefore will
self-dissipate its energy if its size is comparable to that of the
loops, whatever its mass. If the connecting arm indeed traces the
near-side dustlane, the identification of clump~2 with such a
dustlane, as proposed by Stark \& Bania (1986), is ruled out (unless
the Milky~Way has a double bar). But it should be noted that the axis
shock assigned to the connecting arm in model l10't2540 can produce an
$\ell-V$ trace resembling much more that of clump~2 if the bar
inclination angle is reduced to $15\degr$ (see Fig.~\ref{other}).
\par Clump~1, composed of several clouds which are not bound to each
other (Bania et al. 1986), is the southern terminus part of the
135-km\,s$^{-1}$ arm which penetrates the far-side dustlane shock.
According to model~l10't2540, its dynamics should be rather subtle: a
part of its gas is absorbed by the dustlane, resulting in a huge
velocity change like those described above, from
$V\approx 100$~km\,s$^{-1}$ to $V\la 0$, and the other part,
corresponding to the portion of clump~1 at $\ell \la -5\degr$, is
gliding outwards along but without crossing the shock front until
apocentre is reached (see the magenta segment in Fig.~\ref{region}).
The momentum injected by this cloud into the axis shock gas bends the
outer segment of the shock. Since the far-side dustlane is nearly
aligned with the line of sight, the vertical feature near
$\ell=-4\degr$ is in fact a superposition of dustlane gas moving along
this line and of clouds with shock induced velocity gradients (see
Fig.~\ref{path}). Clump~1 must be a very perturbed and compressed
region, and hence a potential site of star formation a priori. An
indicator of ``readiness'' for star formation is given by the
$^{12}$CO $J=2\rightarrow 1$ to $J=1\rightarrow 0$ ratio, tracing
dense molecular clouds, which is indeed particularly high for this
clump (Hasegawa 1997, private communication). However, the strong
shearing in the shock may prevent any star formation to proceed (e.g.
Reynaud \& Downes 1998). The bulk of this clump, owing to its large
mass and impact velocity, may also cross the shock front without being
too much affected.

\subsection{Tilt of the gaseous disc}
%
A further observational argument supporting our interpretation of the
dominant $\ell-V$ features is the fact that the connecting arm and the
portion of the 3-kpc arm at positive longitude, located in the
near-side of the bar according to the models, have their maximum
emission below the Galactic plane (i.e. at $b<0$), and that the
dustlane near $\ell=-4\degr$ and the negative longitude part of the
135-km\,s$^{-1}$ arm, located in the far-side of the bar, above this
plane (i.e. $b>0$; see Fig.~\ref{lvb}). Hence structures predicted to
be spatially close to each other by the models are indeed found at
similar latitude in the observations.
\par Consequently, the gaseous disc within the central $1-2$~kpc is
tilted relative to the plane $b=0$. Referring to Fig.~\ref{lvb},
emission of the connecting arm at $\ell \approx 5\degr$, i.e. where
$b$ stops to decrease as $\ell$ increases, occurs near $b=-1\degr$.
Approximating the dustlanes by a straight line across the Galactic
centre with an in-plane inclination angle of $\varphi_{\circ}=25\degr$
(see Sect.~\ref{con}), the source of this emission is tilted by
$\theta=\tan^{-1}\!|\sin{\varphi_{\circ}}\tan{b}/\sin{\ell}|=4.8\degr$
out of the Galactic plane, corresponding to a distance of about
$120$~pc below this plane. Similarly, the far-side dustlane ends at
$\ell \approx -4.5\degr$ with $b\approx 0.8\degr$, implying a tilt
$\theta=4.3\degr$ and a height of $\sim +135$~pc. A similar tilt is
also apparent on the $^{12}$CO longitude-latitude map of Dame et
al.~(1987; 1999), where the dustlanes contribute only very little to
the total emission. Blitz \& Spergel~(1991) have inferred an apparent
$5.5\degr$ central tilt of the bar major axis from balloon
$2.4$~$\mu$m observations of the Galactic bulge within
$|\ell|<12\degr$ and $|b|<10\degr$. However, such a tilt has not been
confirmed by the more recent \mbox{near-IR} COBE/DIRBE maps (Weiland
et al. 1994). In the dust subtracted K-band map (paper I), the
latitude centroid as a function of longitude, excluding the high
extinction zone $|b|<3\degr$, only shows a significant tilt when
regions beyond $10\degr$ from the Galactic plane are included, but
then the tilt is likely an artifact due to the growing contribution of
zodiacal light. The small tilt angle derived here is not generated by
a position of the Sun above the Galactic plane, which is only of order
$10-40$~pc (Humphreys \& Larsen 1995 and references therein), and
concerns gas on a larger scale than the 180-pc molecular ring, for
which an apparent tilt angle of $\sim 6\degr$ has been estimated, also
with negative latitude in the first Galactic quadrant (Uchida et
al. 1994a; Morris \& Serabyn 1996).
\par Larger tilts of the inner $\sim 2$~kpc gaseous disc, of
$20\degr-30\degr$ and based on expanding or elliptical motion models,
have been reported in the past (e.g. Cohen 1975; Burton \& Liszt 1978;
Liszt \& Burton 1980). More recently, Burton \& Liszt~(1992) have
updated and improved their tilted disc model into a flaring warp with
a central disc coplanar to the Galactic plane. The warp is rectilinear
in each radius and the tilt of its midplane is given by
$\theta_{\rm max}\cdot \sin{(\phi-\phi_{\circ})}$, where $\phi$ is the
galactocentric azimuth relative to the Sun defined positive in the
direction of Galactic rotation, $\phi_{\circ}=45\degr$ the azimuth of
the line of nodes and $\theta_{\rm max}=13\degr$ the maximum tilt
angle. This model predicts a tilt angle of about $4\degr$ at
$\phi=\varphi_{\circ}=25\degr$, in agreement with the tilt derived for
the Galactic dustlanes.
\par The departure of the dustlanes from the plane $b=0$ does not
appear to increase linearly with galactocentric distance, but seems to
stabilise or gently decline at larger distance, rendering its
description by a constant tilt somewhat oversimplistic.

\subsection{Central molecular zone}
\label{bin}
This part of the $\ell-V$ observations is probably the most complex
and the most difficult to understand. Binney et al.~(1991) have given
a detailed description of the gas flow in a rotating barred potential
in terms of the closed $x_1$ orbits, associating the terminal velocity
curves in the observed $\ell-V$ diagrams with the envelope of non
self-intersecting such orbits and the parallelogram structure of the
180-pc molecular ring with the innermost orbit of this kind, called
the ``cusped $x_1$'' orbit, where shocks would transform most of the
atomic gas into molecules and force the gas to plunge onto the more
viable orbits of the $x_2$ family. 
\begin{figure}[t!]
\centerline{\psfig{figure=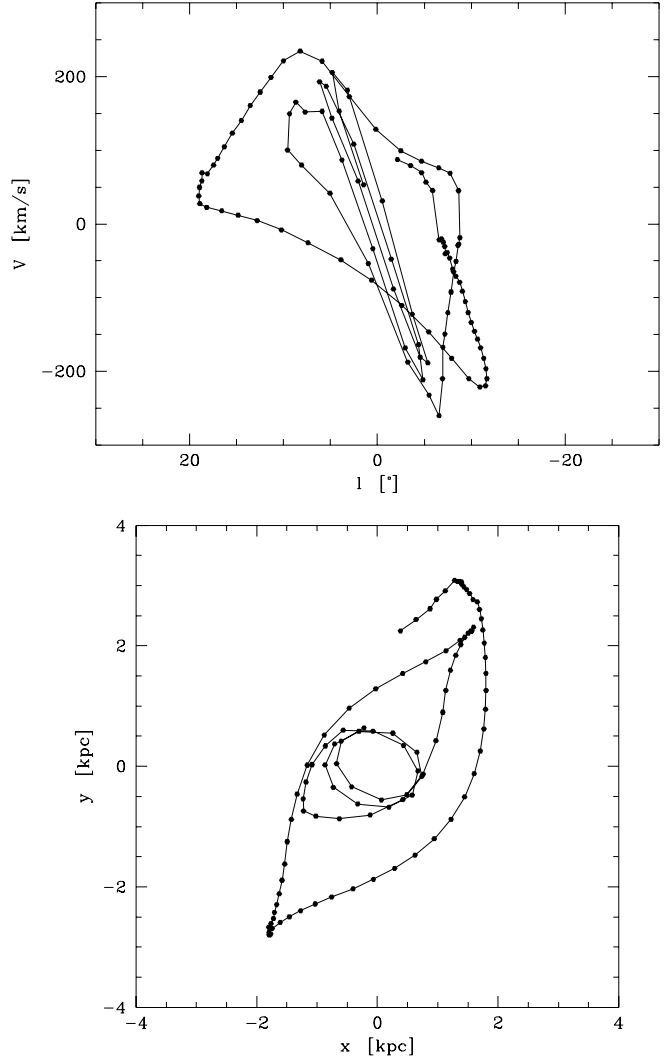,width=8.8cm}}
\caption{Typical path in the $x-y$ and $\ell-V$ planes of an SPH
         particle in simulation l10. The points show the positions of
         the particle from $t=1994$ to $2200$~Myr with a constant time
         interval of $2$ Myr. The $x-y$ plot is in the bar rotating
         frame with the origin at the density centre and with the bar
         inclined by $25\degr$ relative to $x=0$ (like in
         Fig.~\ref{models}), and the $\ell-V$ plot is viewed from
         $(x,y)=(0,-8)$~kpc. The particle is chosen to pass through
         the intersection of the near-side lateral arm and axis shock
         at $t=2066$~Myr, i.e. through $(x,y)=(-1.4,-2.4)$~kpc in the
         lower left frame of Fig.~\ref{models}. Note the larger
         velocity gap between the points when the particle crosses the
         far-side axis shock.}
\label{path}
\end{figure}
\begin{figure*}[t!]
\centerline{\psfig{figure=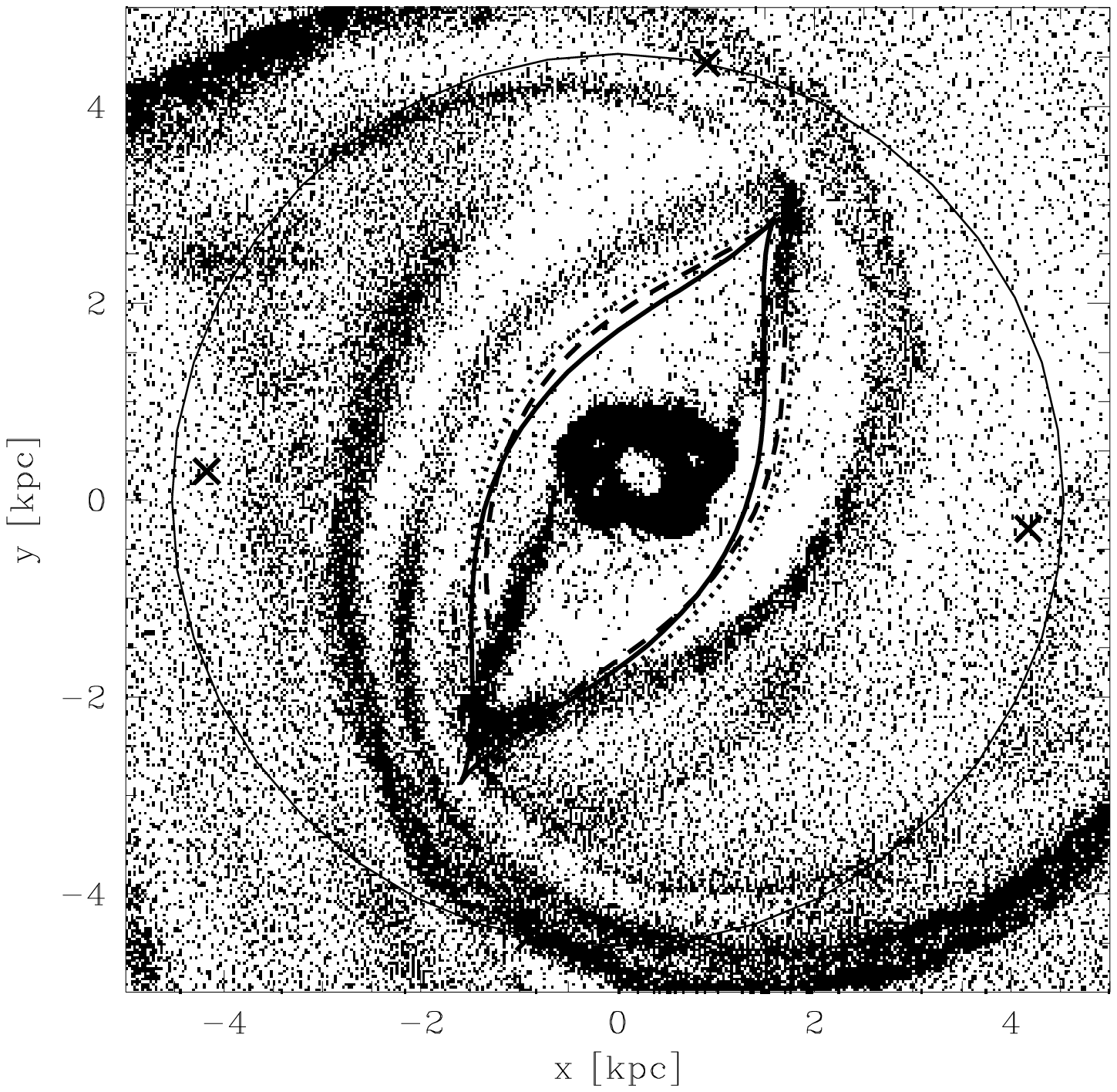,height=7cm}
            \psfig{figure=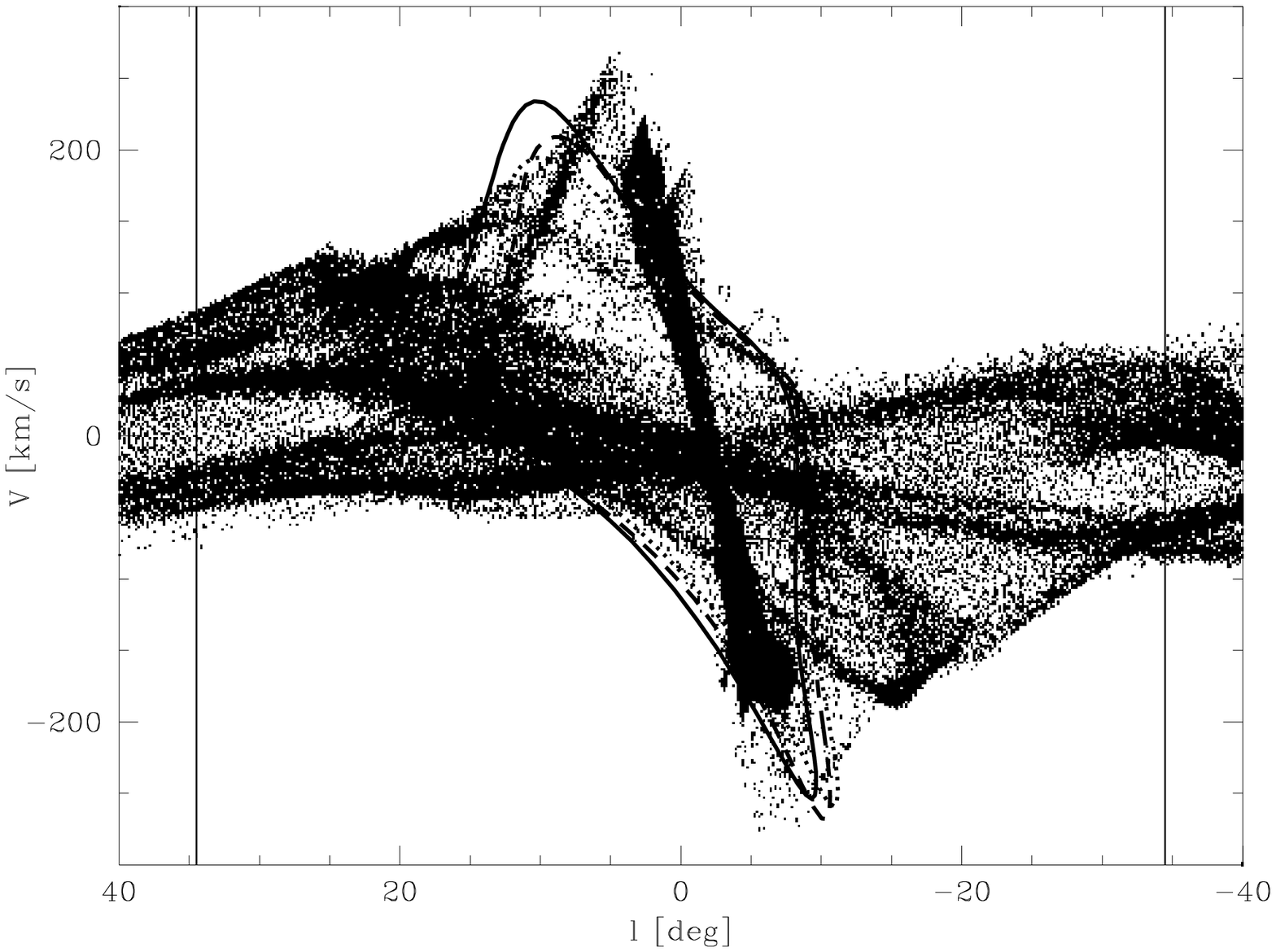,height=7cm}}
\caption{Cusped $x_1$ orbits in the instantaneous frozen rotating
         potential, superposed on the spatial gas distribution of
         model~l10t2066 (left),~and the corresponding traces in the
         $\ell-V$ diagram (right). The solid line orbit is obtained
         imposing reflection symmetry on the potential with respect to
         the vertical axis through the centre of mass, and the other
         orbits are based on the exact potential rotating about the
         centre of mass, the dashed and dotted orbits having single
         cusps at opposite ends. The crosses indicate the Lagrangian
         points $L_1$, $L_2$, $L_4$ and $L_5$, and the thin solid
         lines the corotation circle, based on the radius of the
         Lagrangian points $L_1$ and $L_2$, and the longitudes of its
         tangent points. Note that in the particle representation used
         here, the $\ell-V$ plot looks different from the one in
         Fig.~\ref{models} because the weights of the inverse squared
         distance relative to the observer are not taken into account.
         For example, the molecular ring is more difficult to
         distinguish without this correction.}
\label{x1}
\end{figure*}
\par This model is confronted to several problems. The most frequently
reported one (e.g. Kuijken 1996; Morris \& Serabyn 1996) is the
asymmetry of the observed 180-pc parallelogram, and in particular the
substantial $\sim 140$~km\,s$^{-1}$ forbidden velocity near
$\ell=-0.8\degr$, which cannot be fully accounted for by projection
effects of the cusped $x_1$ orbit. Other problems are: (i) when viewed
in the full $\ell-b-V$ data cube, the 180-pc parallelogram appears
more as an assemblage of larger scale features rather than forming a
distinct unity (see e.g. Fig.~4 in Morris \& Serabyn 1996); for
instance, its upper edge is observed to extend continuously well
beyond the longitude range of the postulated parallelogram and joins
the connecting arm through the positive terminal velocity peak,
(ii)~the terminal velocity peaks and the parallelogram cannot both be
generated by the same cusped $x_1$ orbit because the formers occur
well outside the longitude boundaries of the latter, (iii) the
longitudinal edges of the parallelogram near $\ell=-1\degr$ and
$1.5\degr$, assimilated to the bar leading shocks in the model, should
define rather clear $\ell-V$ features as in our models, yet these
edges are more disordered than the other edges of the parallelogram,
and (iv) the cusped $x_1$ orbit of the stellar dynamical models in
paper~I generally does not match the HI terminal velocity peaks; the
models providing the best agreement always arise in young and not
completely stabilised bars, whereas in older bars the cusped $x_1$
orbit presents velocity peaks at fairly larger absolute longitudes
than observed.
\par Figure~\ref{x1} shows that the dustlane shocks are indeed
responsible for the positive and negative terminal velocity peaks, but
that they do not coincide with the leading edges of the cusped
$x_1$~orbit, which produces velocity peaks at higher absolute
longitude and with lower velocity amplitude. Gas on the shocks move
along non-periodic orbits with much smaller pericentre than the cusped
$x_1$~orbit. Thus the Milky~Way's cusped $x_1$~orbit is probably much
larger than in the Binney et al. model, explaining why these authors
find a very small corotation radius of $2.4$~kpc. The asymmetries in
our models make the cusped $x_1$ orbit slightly uncertain. In
particular, there is a small range of Hamiltonians where the $x_1$
orbits only loop at one extremity. However, studies of Galactic $x_1$
orbits generally represent the true Milky~Way's potential by
bisymmetric models and therefore are also concerned with similar
uncertainties.
\par The gas on the dustlanes falling onto the nuclear ring from large
distances does not necessarily merge with the ring, but rather passes
round of it and lands on the opposite dustlane closer to the nuclear
ring, merging with it only at the next passage or after repeating the
whole cycle once more (Fig.~\ref{path}; see also Fig.~10a in Fukuda et
al. 1998). The farther out the gas on the dustlanes originates, the
more it passes away from the nuclear ring. The upper and lower edges
of the 180-pc parallelogram could represent such gas streams which
precisely brush the nuclear ring/disc, loosing only little mass to it.
The mass transfer would appear in the $\ell-V$ plot like vertical
bridges between the streams and the nuclear disc. An example of such
bridges is detectable in the high resolution $^{12}$CO data (e.g.
Morris \& Serabyn 1996) near $\ell=1.3\degr$ and for $V$ between~$100$
and $200$~km\,s$^{-1}$. The brushing streams finally crash in the
dustlanes where they are abruptly decelerated. The right longitudinal
edge of the 180-pc parallelogram, at $\ell\approx -0.8\degr$, could be
a trace of this process. However it is not very clear why this edge is
located at lower absolute longitude than the bright emission near the
negative terminal velocity peak, corresponding to gas with about
maximum velocity on the far-side dustlane, i.e. why the gas on this
dustlane slows down {\it before} being striked by the positive
velocity stream round the nuclear ring. Uchida et al.~(1994b) have
reported on a large scale shock front in the AFGL~5376 region, which
is close to the southern end of the parallelogram upper edge, but
which they consider as a support of the expanding molecular ring
hypothesis.
\begin{figure*}[t!]
\centerline{\psfig{figure=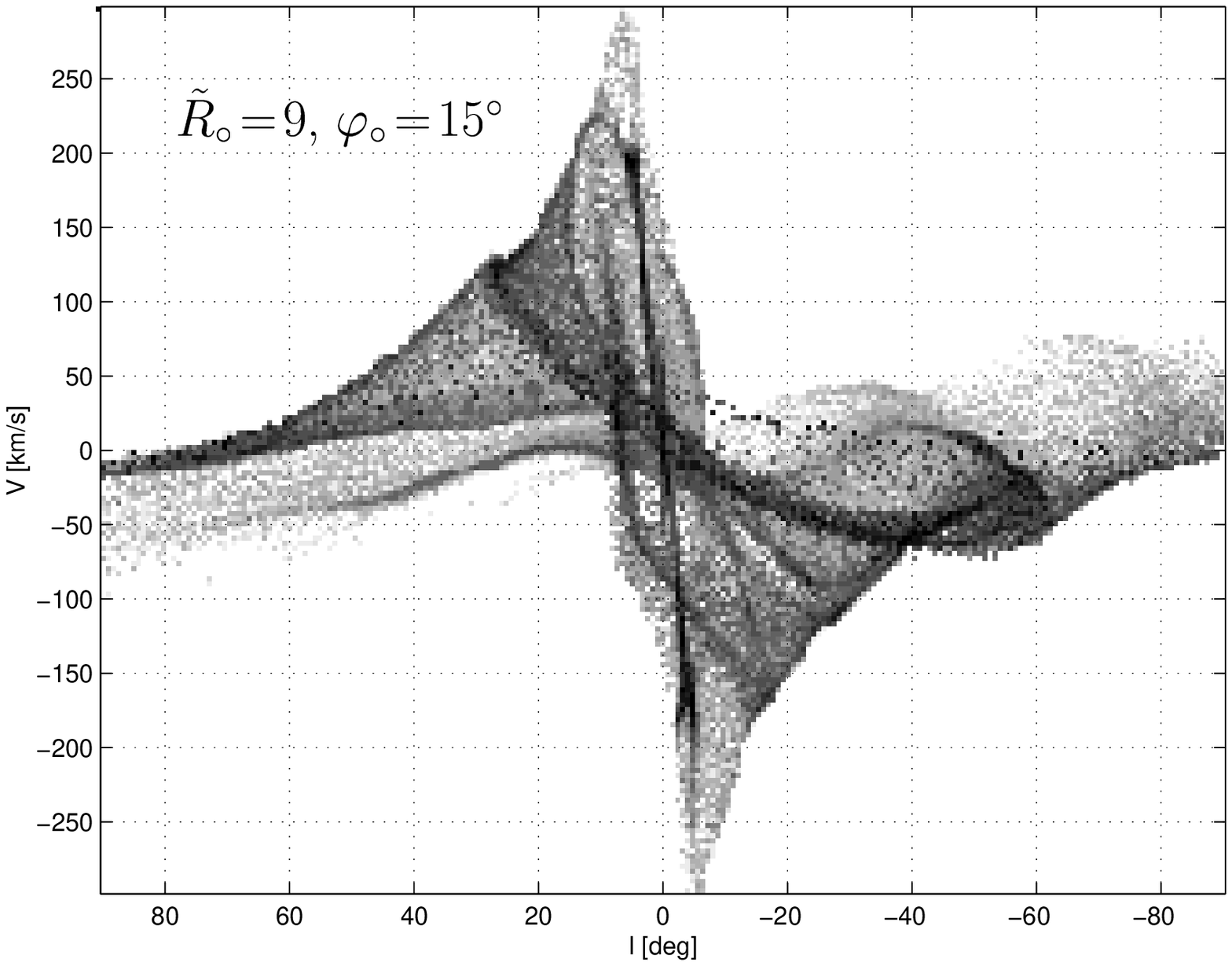,width=8.75cm}
            \psfig{figure=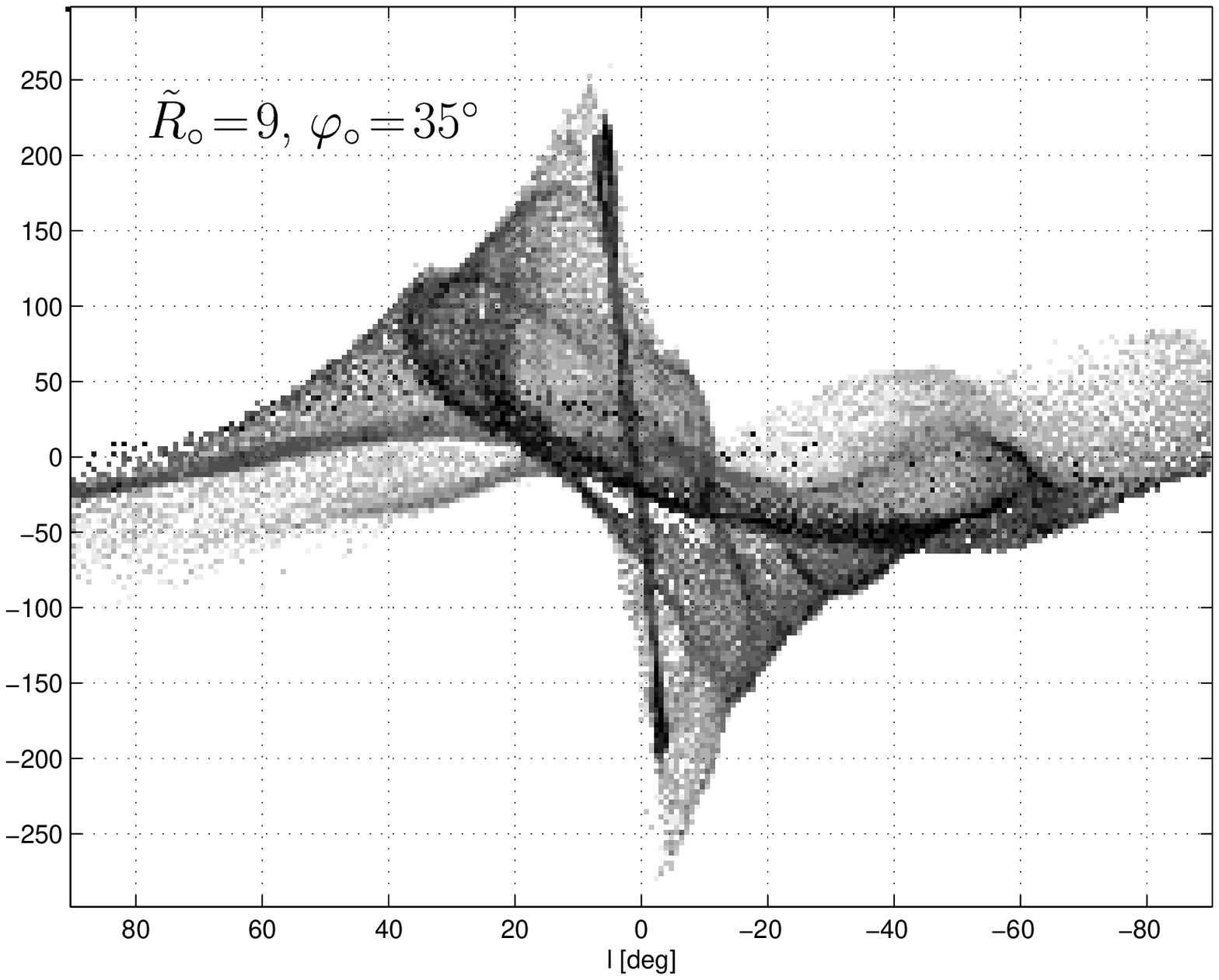,width=8.75cm}}
\centerline{\psfig{figure=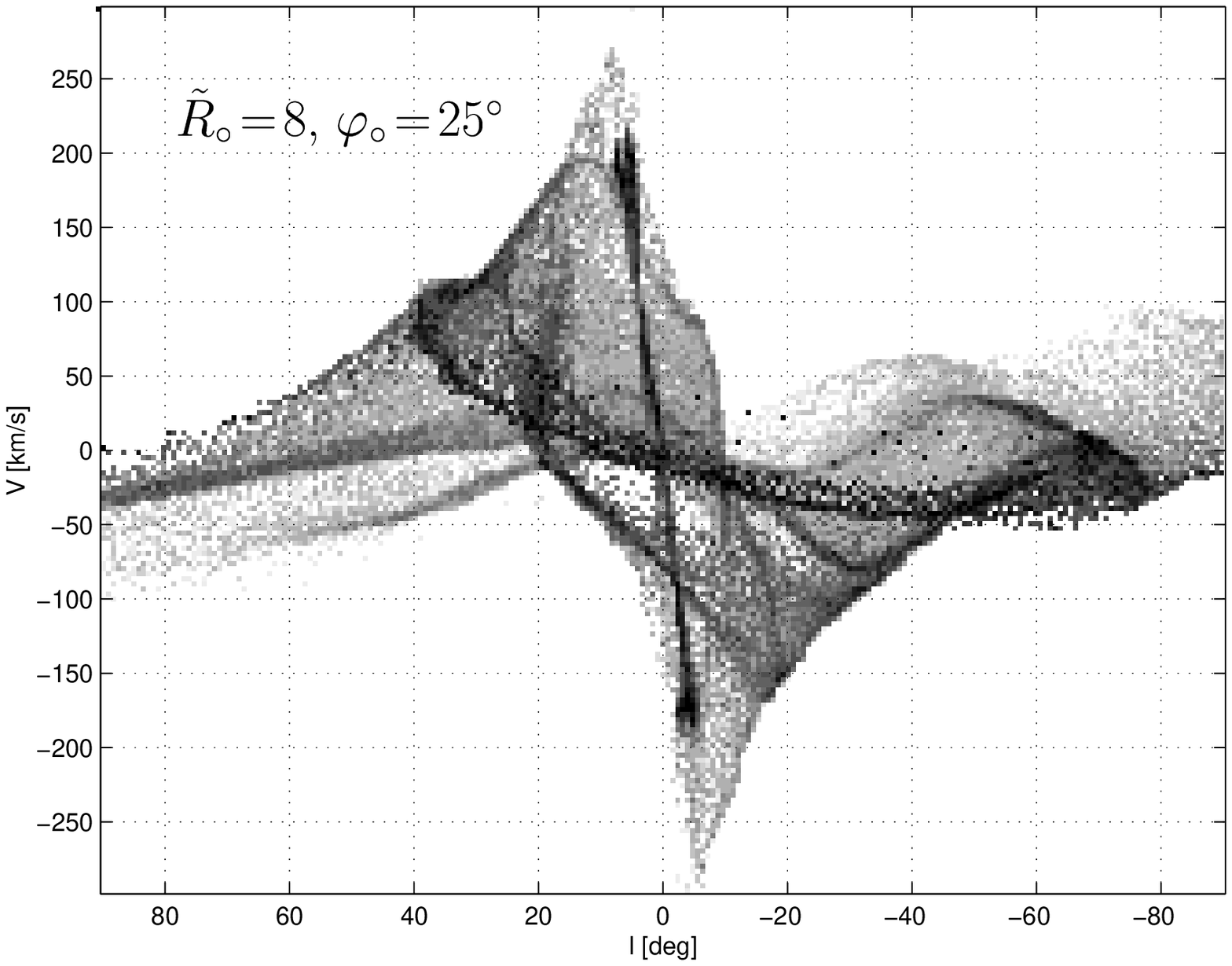,width=8.75cm}
            \psfig{figure=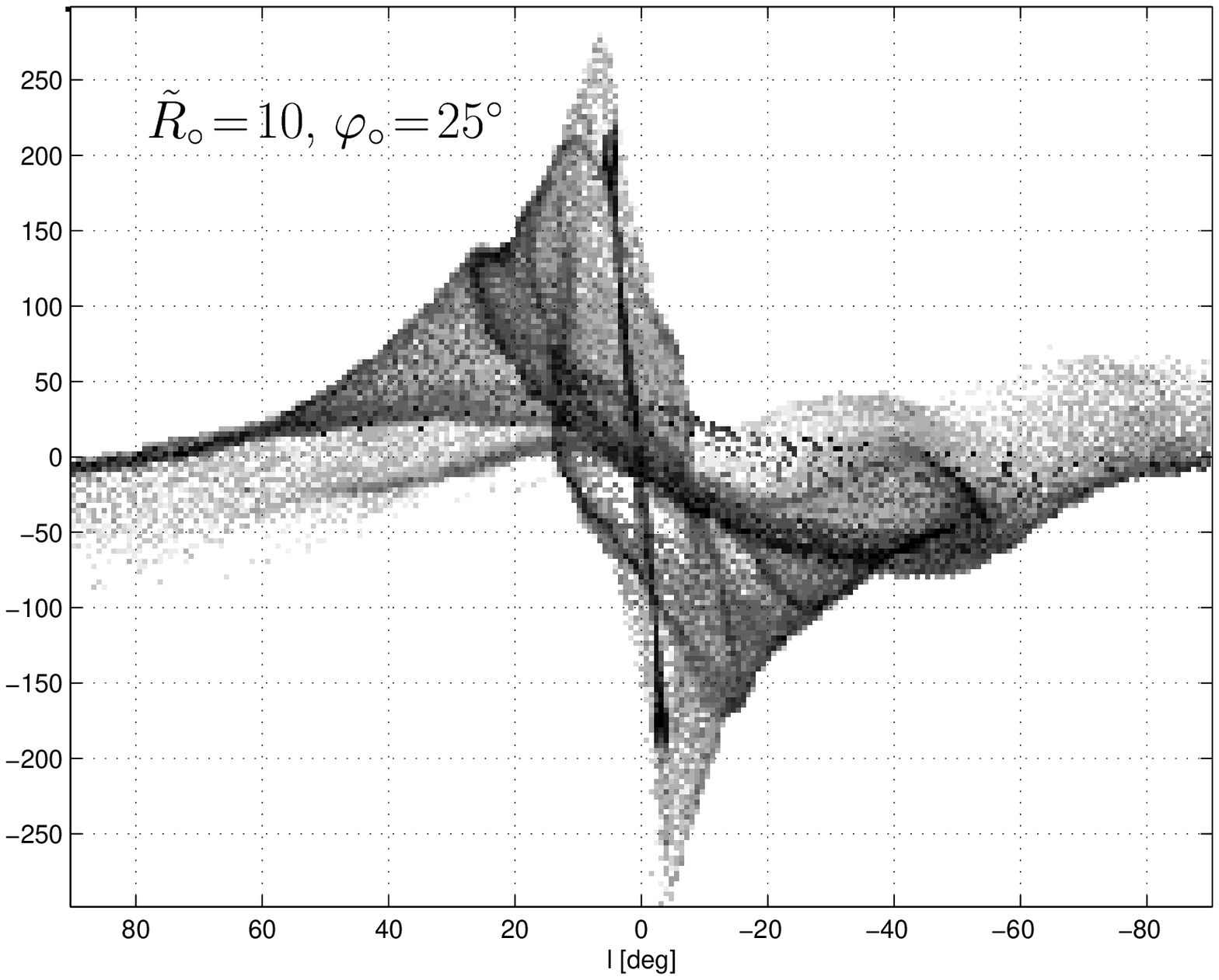,width=8.75cm}}
\caption{Dependence of the $\ell-V$ diagrams on the location of the
         observer relative to the bar in model l10't2540. The upper
         diagrams illustrate the effects of changing the bar
         inclination angle and the lower ones the effects of distance.
         $\tilde{R}_{\circ}$ represents the galactocentric distance of
         the observer in initial units, which is $9$ in the adopted
         optimum model displayed in Fig.~\ref{models}.}
\label{other}
\end{figure*}

\section{Geometrical constraints on the bar parameters}
\label{con}
Figure~\ref{other} shows how the $\ell-V$ diagram of model l10't2540
changes when the viewing point of the observer is modified. Reducing
the bar inclination angle $\varphi_{\circ}$ shrinks the structures
longitudinally and amplifies the velocities. The observed knee of the
135-km\,$s^{-1}$ arm near $\ell=-5\degr$ is better reproduced, but the
transition of the 3-kpc arm to the connecting arm happens at too
negative velocity and the connecting arm becomes too steep. Increasing
the angle $\varphi_{\circ}$ lowers the forbidden velocities, moves the
terminal velocity peaks further away from $\ell=0$ and shifts the
connecting arm closer to the northern tangent points of the two
molecular ring branches. Increasing the distance of the observer
obviously produces a longitudinal contraction, but without modifying
the velocity of the structures near the centre. The diagram with the
observer at $\tilde{R}_{\circ}=8$ moves the loop associated to the
$\ell<0$ prolongation of the molecular ring structure out to the real
tangent point of the Carina arm near $\ell=-78\degr$. These properties
cannot be used to infer a robust inclination angle of the bar because
they are based on one specific model and the gas flow is strongly
time-dependent.
\par
\begin{figure}[t!]
\centerline{\psfig{figure=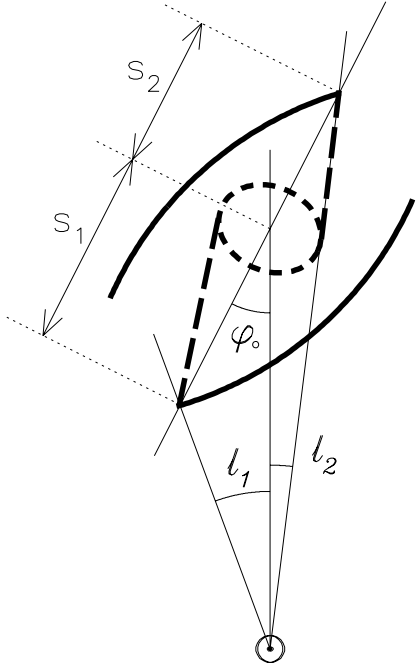,width=7cm}}
\caption{Basic geometry for the localisation of the intersections
         between the lateral arms (solid lines) and the axis shocks
         (long dashed lines). The nuclear ring (short dashed line) has
         been enlarged and the location of the Sun is indicated by the
         $\odot$ symbol.}
\label{angle}
\end{figure}

 Constraining the bar parameters by adjusting gas dynamical models
to the observed CO and HI $\ell-V$ diagrams is a very delicate task
and may lead to unreliable results if the models are not sufficiently
realistic. However, with our interpretation of the dominant features
in these diagrams, it is possible to provide geometrical constraints
on the bar inclination angle and extension which do not depend on the
details of the models. The principle of the method, illustrated in
Fig.~\ref{angle}, is to determine in the CO and HI data the longitudes
$\ell_1$ and $\ell_2$ where the 3-kpc and the 135-km\,s$^{-1}$ arms
intersect the axis shocks and to adjust in real space a major axis
through the Galactic centre crossing the line of sights associated to
these two directions at galactocentric distances in the ratio
$q\equiv s_1/s_2$. Simple trigonometrical considerations in the first
and fourth Galactic quadrants respectively yield:
\renewcommand{\theequation}{\arabic{equation}-27}
\begin{equation}
\frac{s_i}{R_{\circ}}=\frac{\sin{|\ell_i|}}{\sin{(\varphi_{\circ}+\ell_i)}},
                        \hspace{.5cm}i=1,\:2,
\end{equation}
\renewcommand{\theequation}{\arabic{equation}}
\addtocounter{equation}{1}
and isolating $\varphi_{\circ}$ and $s_1$ between these two equations:
\begin{eqnarray}
\tan{\varphi_{\circ}} & = & -\frac{1+q}{\cot{\ell_2}+q\cot{\ell_1}},
                                                       \label{phio} \\
\frac{s_1}{R_{\circ}}    & = &
                  \left[(1+q)\frac{\sin^2{\ell_1}+q\sin^2{\ell_2}}
                  {\sin^2{(\ell_1-\ell_2)}}-q\right]^{1/2},
\label{aro}
\end{eqnarray}
where $q$ remains as a parametrisation of the asymmetry level between
the lateral arms, which is not a priori known. In the ideal
bisymmetric case $q=1$, but in reality $q>1$ according to
Sect.~\ref{lat}. These formulae rest on the implicit assumption that
the intersections of the lateral arms with the axis shocks and the
Galactic centre are collinear. If this is wrong, then
$\varphi_{\circ}$ will represent the angle between the line joining
these intersections and the direction $\ell=0$. 
\par
\begin{figure}[t!]
\centerline{\psfig{figure=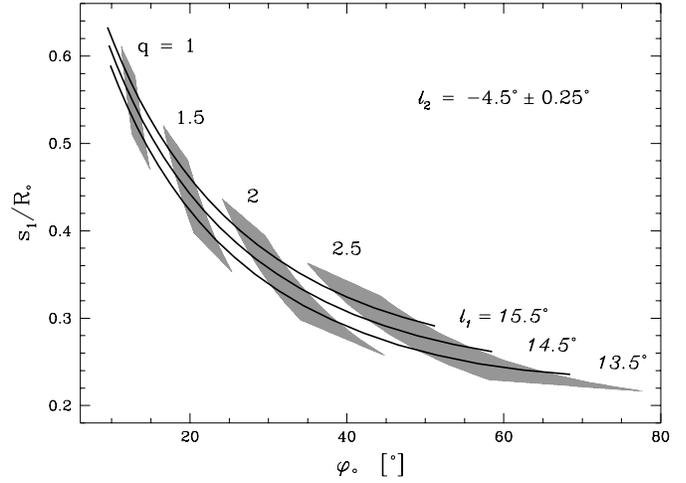,width=8.8cm}}
\caption{Graphical representation of the geometrical constraints on
         the bar parameters. The solid curves are derived from
         Eq.~(26) for different values of $\ell_1$, and the grey
         surfaces from Eq.~(27) for different values of $q$, setting
         $s_2=s_1/q$ and assuming
         $-4.75\degr\leq \ell_2\leq -4.25\degr$ to show how
         uncertainties in $\ell_2$ propagate to $q$.}
\label{phi}
\end{figure}

 Extrapolating the connecting arm down to the 3-kpc arm in the
$\ell-V$ observations leads to $\ell_1\approx 14.5\degr$, where a knee
of the 3-kpc arm can be barely detected, whereas the 135-km\,s$^{-1}$
arm meets the opposite axis shock at $\ell_2\approx -4.5\degr$
(Fig.~\ref{lv}). The resulting constraints are put together in
Fig.~\ref{phi}. Clearly, the distance $s_1$ increases for smaller
values of $\varphi_{\circ}$, but no precise value can be given so far
for the inclination angle. However, the CO data in Fig.~\ref{lv}
betray a second fainter far-side lateral arm which extends down to
longitude $\ell_2= -7\degr\pm 0.5\degr$ with lower forbidden
velocities than the 135-km\,s$^{-1}$ arm, as well as a quasi-vertical
feature at about the same longitude probably corresponding to gas from
the same arm plunging towards the nuclear ring/disc after apocentre,
on orbits parallel to the main axis shock. If this is correct, this
arm must be much more symmetrical to the 3-kpc arm and the
formulae~(\ref{phio}) and~(\ref{aro}) can be applied to these two arms
using $q=1$. The results are $\varphi_{\circ}=25\degr\pm 4\degr$ and
$s_1=3.15$~kpc, and by the way $q=1.76$, i.e. $s_2=1.8$~kpc, for the
135-km\,s$^{-1}$ arm.
\par It remains to see how the arm intersections discussed here are
related to the true bar parameters. Numerical simulations and analyses
of observations in early-type barred galaxies indicate that the ratio
between the bar semi-major axis and the corotation radius amounts to
$a/R_L=0.85\pm 0.15$, and offset dustlanes in such galaxies do not
extend beyond the bar ends, i.e. $s_i\leq a$ (e.g. the review of
Elmegreen 1996). Also, bisymmetric hydro simulations in rotating
barred potentials with straight offset dustlanes generally have
lateral arms intersecting the axis shocks very close to their outer
ends and at most a few degrees ahead of the bar major axis. In the
standard model 001 of Athanassoula~(1992), which has an inner Lindblad
resonance but no looped $x_1$ orbits, $s/R_L\approx 0.72$. Adopting
$s_1/R_L=0.8\pm 0.1$ as a good compromise, our value of $s_1$ would
imply a corotation radius of $4.0\pm 0.5$~kpc for the Milky~Way.
\par
\begin{figure*}[t!]
\centerline{\psfig{figure=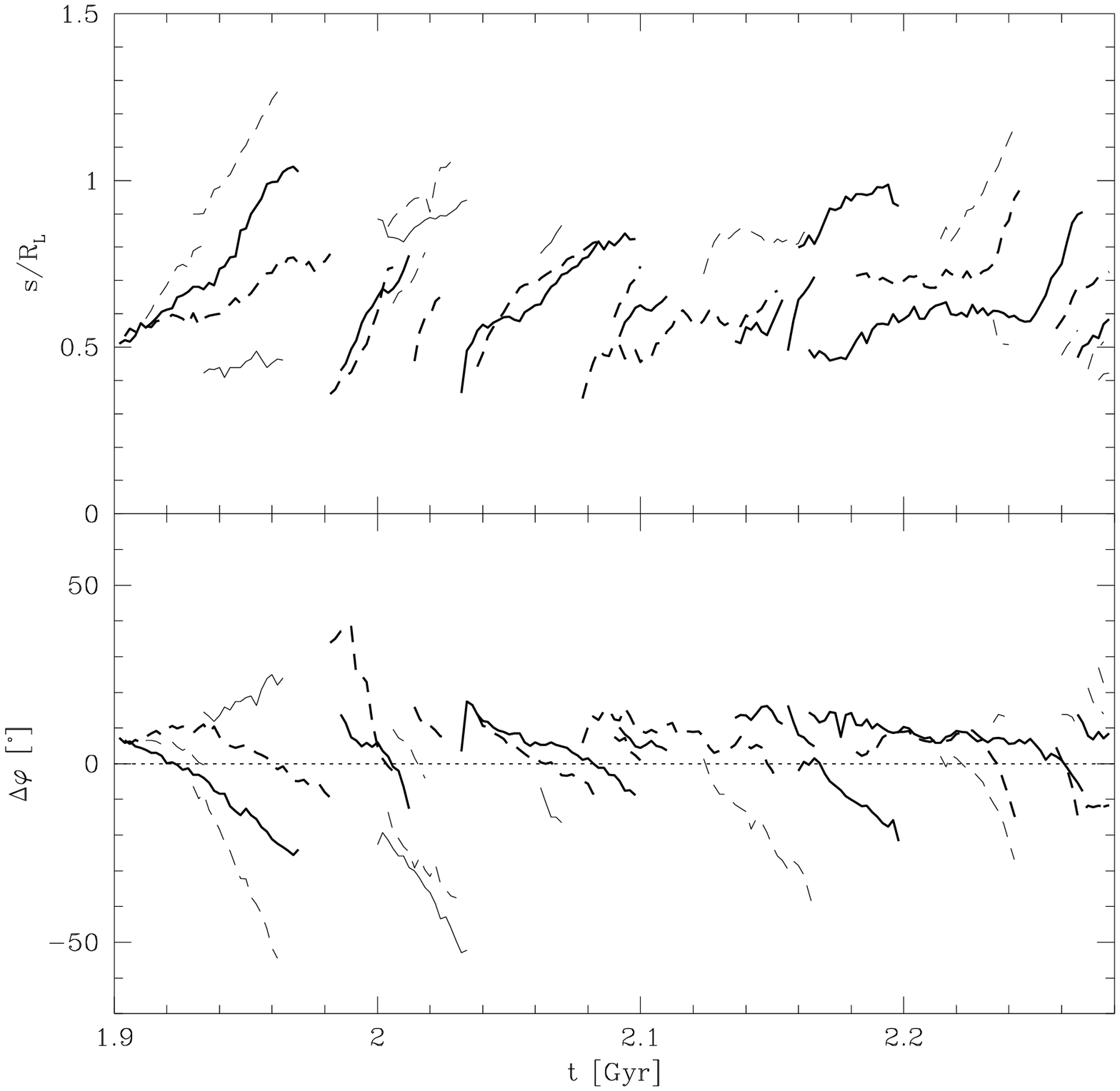,height=10.2cm}\hspace*{.5cm}
            \psfig{figure=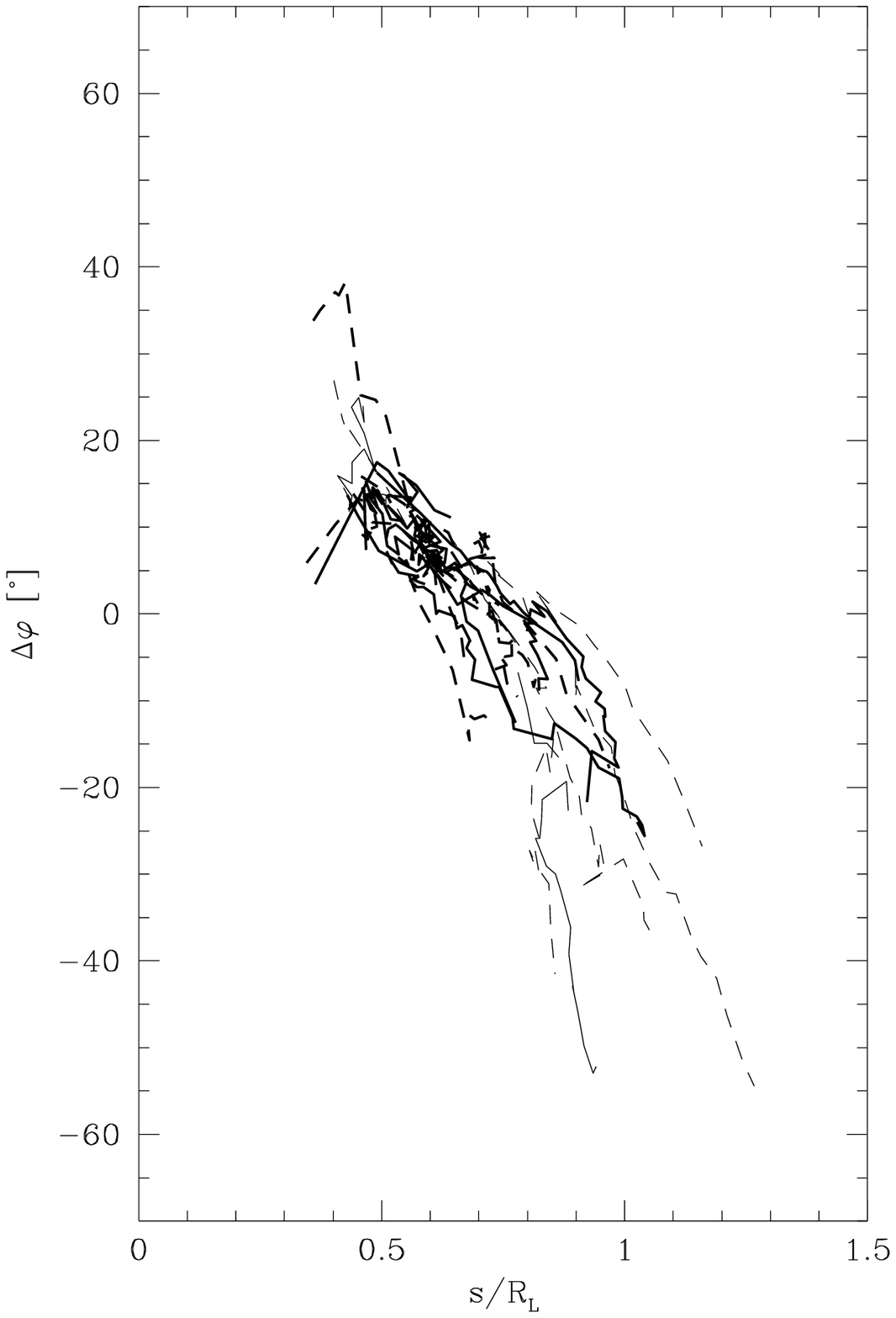,height=10.2cm}}
\caption{Location of the intersections between the lateral arms and
         the axis shocks relative to the bar in simulation l10. Left
         top: distance of these intersections relative to the
         offcentred density centre and normalised by the corotation
         radius (based on the Lagrangian points $L_1$ and $L_2$). Left
         bottom: phase of the intersections relative to the major axis
         of the bar, defined positive when the formers lead the
         latter; the dotted line indicates the locus where both are
         exactly in phase. Right: correlation between the ordinates of
         the two previous plots. The solid and dashed lines refer to
         arm intersections on the same side of the bar, and the bold
         lines to intersections involving the most contrasted lateral
         arms.}
\label{lateral}
\end{figure*}

 The situation in our non-symmetrised and time-dependent
simulations, shown in Fig.~\ref{lateral}, is much more complicated
however. As mentioned in Sect.~\ref{gas}, the radius of the
intersections between the lateral arms and the axis shocks increases
with time. At the formation of a lateral arm, the associated
intersection leads the bar major axis, and as the arm moves outwards,
it crosses this axis and becomes trailing. When both are exactly in
phase, the intersection radius relative to corotation is $0.75$ on the
average, compatible with the value adopted above, and compares very
well to the apocentre radius of the cusped $x_1$ orbit (see
Fig.~\ref{x1}). Further out, the lateral arms rapidly dissolve in the
spiral arms emanating from the very end of the axis shocks. The
minimum value of the ratio $s_i/R_L$, which could depend on the size
of the nuclear ring in the models, is about $0.4$. Taking this value
as a lower limit for the 135-km\,s$^{-1}$ arm leads to
$s_1/R_L=q\cdot s_2/R_L\ga 0.7$ for the 3-kpc arm, suggesting that
this arm should meet the connecting arm close to the bar major axis
and close to the apocentre of the cusped $x_1$ orbit. This is also
confirmed by the fact that there is no obvious velocity gap in the
observed $\ell-V$ diagrams at the longitude where the 3-kpc arm passes
into the connecting arm. Our models suggest that the tangent points of
the two molecular ring branches in the first Galactic quadrant are
within corotation (see Fig.~\ref{x1}), contrary to Englmaier \&
Gerhard's~(1999) deduction, and therefore we rather advocate
$R_L\geq 4$~kpc.
\par Our value for the bar inclination angle agrees very well with
many other determinations (e.g. Stanek et al. 1997; Nikolaev \&
Weinberg 1997), and in particular with the recent result
$\varphi_{\circ}=20\degr-25\degr$ of Englmaier \& Gerhard~(1999), who
derived the gas flow in the potential of the deprojected COBE/DIRBE
\mbox{L-band} luminosity distribution (Binney et al. 1997). The
bisymmetric hydro simulations done by Weiner \& Sellwood~(1999) and
matched to the observed HI terminal velocity curves favour slightly
larger values of this angle, i.e. $\varphi_{\circ}=35\degr\pm 5\degr$,
and support a corotation radius between $4$ and $6$~kpc (for
$R_{\circ}=8.5$~kpc). These authors consider an angle below $25\degr$
unlikely, but it should be noted that gas flow fluctuations like those
in our self-consistent simulations can increase the $\ell-V$ area
covered by forbidden velocities and hence substantially bias their
method.
\par It is also possible from our data interpretation to give a crude
estimate of the bar pattern speed. If the 3-kpc arm indeed encounters
the connecting arm near the apocentre of the cusped $x_1$~orbit, then
the gas at the intersection of the two arms should corotate with the
bar figure. Evaluating the radial velocity~$\Delta V$ of this gas
relative to the LSR yields its rotation velocity with respect to the
Galactic centre:
\begin{equation}
V_{\rm c}(s_1)=\frac{\Delta V+V_{\circ}\sin{\ell_1}}
                    {\sin{(\ell_1+\varphi_{\circ})}},
\end{equation}
and resorting to Eq.~(26):
\begin{equation}
\Omega_{\rm P}\equiv \frac{V_{\rm c}(s_1)}{s_1}=\frac{\Delta V}
                          {R_{\circ}\sin{\ell_1}}+\Omega_{\circ},
\end{equation}
where $V_{\circ}$ is the circular velocity of the LSR and
$\Omega_{\circ}=V_{\circ}/R_{\circ}$. The bar inclination angle and
$s_1$ have vanished in this last formula and thus the method is
independent of them. The value of $\Delta V$ is hard to determine in
the CO and HI $\ell-V$ plots because of overlayed emission from the
disc and in particular from the molecular ring, but a reasonable range
is $\Delta V=15-35$~km\,s$^{-1}$. With $R_{\circ}=8$~kpc,
$V_{\circ}=220$~km\,s$^{-1}$ and $\ell_1=14.5\degr$ as before, one
gets $\Omega_{\rm P}=35-45$~km\,s$^{-1}$\,kpc$^{-1}$, in fair
agreement with the $50$~km\,s$^{-1}$\,kpc$^{-1}$ of our gas
simulations\linebreak (Fig.~\ref{omp}) when rescaled to
$R_L=4.25$~kpc. Changing the value of $\ell_1$ or $V_{\circ}$ by
$1\degr$ and $10$~km\,s$^{-1}$ modifies the result only by $\sim 2$\%
and $3$\% respectively. However, $\Delta V$ may be affected by a
radial motion of the LSR and/or oscillations of the bar density
centre. Longitude-velocity maps of very dense gas (like CS) could help
to localise more precisely the transition from the 3-kpc arm to the
connecting arm. Note that the method does not apply to the very
asymmetric model l10't2540, as can be checked from Fig.~\ref{other},
because the gas at the transition is not at rest in the frame of the
bar.

\section{Conclusion}
%
This paper presents the first fully self-consistent gas flow models of
the inner Milky~Way, obtained by incorporating an SPH component with
$150\,000$ particles in realistic symmetry-free barred 3D $N$-body
simulations with nearly $4\times 10^6$ total number of particles. The
axisymmetric initial conditions of the simulations are chosen to
evolve spontaneously in a barred configuration compatible with the
COBE/DIRBE K-band constraints, according to criteria based on a set of
lower resolution pure stellar dynamical simulations realised in a
precedent paper. The stellar bar rotates with its intrinsic natural
pattern speed and the gas component, gently released from its
axisymmetric configuration after the formation of the bar, freely
interacts with the live stellar arms.
\par The density centre of the stellar bar becomes unstable and
oscillates around the centre of mass with an amplitude of several
$100$~pc and a frequency of $20-30$~km\,s$^{-1}$\,kpc$^{-1}$. The gas
flow, contrary to other bisymmetric hydro simulations in fixed
rotating barred potentials, never reaches a quasi-steady state and can
delineate strong asymmetries in its spiral structure, not necessarily
induced by the lopsided stellar distribution.
\par Some models selected from the simulations account for many
features seen in the HI and CO longitude-velocity distributions within
the Galactic bar and surrounding regions, and provide a very powerful
guide to understand the inner structure of the Milky~Way. The Galactic
bar is inclined by $25\degr\pm 4\degr$ relative to the $\ell=0$ line,
has a corotation radius of $4.0-4.5$~kpc, a related pattern speed
$\Omega_{\rm P}\sim 50$~km\,s$^{-1}$\,kpc$^{-1}$, and a face-on axis
ratio $b/a\approx 0.6$. As in most early-type barred spirals, offset
dustlanes are leading the bar major axis. Their gaseous traces in the
observed $\ell-V$ diagrams correspond to the connecting arm (near-side
branch) and another feature near $\ell=-4\degr$ (far-side branch).
These dustlanes are the loci of strong shearing shocks with velocity
changes up to $200$~km\,s$^{-1}$ across, and are located closer to the
bar major axis than the leading edges of the cusped $x_1$ orbit. The
peaks in the terminal velocity curves at $\ell\approx \pm 2.5\degr$
are produced by the post-shock gas and not by the $\ell-V$ trace of
the latter orbit, which has velocity peaks at larger absolute
longitude and with lower velocity amplitude. The near-side branch of
the dustlanes lies below the Galactic plane ($b<0$) and the other
branch, which is seen almost end-on, above it. Their maximum departure
fromthe plane amounts to more than $100$~pc.
\par The 3-kpc and 135-km\,s$^{-1}$ arms are the inner prolongations
of spiral arms in the disc, each passing close to one bar end and
joining by a large bow around the nuclear ring/disc the dustlane in
the other side of the bar, at galactocentric distances
$R\approx 3.2$~kpc and $1.8$~kpc respectively. The 135-km\,s$^{-1}$
arm is associated with larger forbidden velocities because of its
deeper location in the central potential well.
\par Velocity-elongated features in the observed CO $\ell-V$ diagrams
``below'' the connecting arm, i.e. at similar longitudes but lower
velocities than this arm, are interpreted as gas lumps which are just
about to cross the shock front layer of the near-side dustlane. A
robust example of such a feature is given by the vertical feature near
$\ell=5.5\degr$. Bania's clump~2 complex is a more controversial
candidate because its mean Galactic latitude differs by roughly half a
degree from that of the connecting arm at same longitude and because
of its substantial mass. Finally, Bania's clump~1 complex is the part
of the 135-km\,s$^{-1}$ arm which strikes the far-side dustlane with a
considerable incident velocity of order $100$~km\,s$^{-1}$. A fraction
of its gas is directly absorbed by the dustlane, producing the
apparent positive velocity trace of the dustlane, another fraction is
gliding outwards along the dustlane, but the bulk of the clump is
probably not intercepted by the dustlane.
\par The interpretation of the main observed $\ell-V$ features given
in this paper are by far the most plausible that can be deduced from
our simulations. However, these simulations may still significantly
depart from reality and hence the unicity of our interpretation
cannot be guaranteed. For instance, if a large amount of gas is
concentrated in the centre, a nuclear fast rotating bar may
temporarily form and lead to a different understanding of the very
complex innermost gas dynamics.
\par Some mpeg movies of the gas flow in simulation~l10,\linebreak
including live $\ell-V$~diagrams, are available on the web at 
http://obswww.unige.ch/\,\~{}fux.

\begin{acknowledgements}
I would like to thank L.~Martinet, D.~Pfenniger and D.~Friedli for
many advices and careful reading of the manuscript, D.~Friedli for
making its original hybrid $N$-body and SPH code available, and
F.~Combes (referee of this paper), J.~Sellwood, O.~Gerhard and
P.~Englmaier for several enlightening discussions. I am also very
thankful to T.M.~Dame for providing the new high resolution $\ell-V$
$^{12}$CO data, as well as to B.W.~Burton, S.~West, H.~Liszt, S.~Digel
and J.~Ohlmacher for their help in reassembling the necessary
HI~surveys for a full $\ell-b-V$ coverage of the Galactic plane
region. This work was mainly supported by the Swiss National Science
Foundation.
\end{acknowledgements}

\end{document}